\documentclass[a4paper,twocolumn,11pt,unpublished]{quantumarticle}
\pdfoutput=1
\usepackage[utf8]{inputenc}
\usepackage[english]{babel}
\usepackage[T1]{fontenc}
\usepackage{amsmath}
\usepackage[colorlinks=true,allcolors=violet]{hyperref}

\usepackage{tikz}
\usepackage{lipsum}
\usepackage{upgreek}

\usepackage{array}
\usepackage[numbers,sort&compress]{natbib}

\begin{document}

\title{Symmetry-protected states of interacting qubits in superconducting quantum circuits}

\author{Yi Shi${}^{1}$, Eran Ginossar${}^{2}$, Michael Stern${}^{3}$, Marzena Szymanska${}^{1}$}

\affiliation{${}^{1}$Department of Physics and Astronomy, University College London, London, WC1E 6BT, UK\\
${}^{2}$Department of Physics and Advanced Technology Institute, University of Surrey, Guildford, GU2 7XH, UK\\
${}^{3}$Department of Physics, Technion-Israel Institute of Technology, Haifa 32000, Israel}

\maketitle
\begin{abstract}
    Superconducting circuits are one of the leading candidates for storing and manipulating quantum information. Among them, qubits embedded with intrinsic noise protection have seen rapid advancements in recent years. This noise protection is typically realized by isolating the computational states from local sources of noise. Here, we propose an interacting spin model that requires at least four spins with nearest-neighbor and next-nearest-neighbor couplings, where the two lowest eigenstates form a symmetry-protected qubit manifold, which is robust to both relaxation and dephasing from local perturbations. We map the spin model to a superconducting circuit and show that such a circuit can reach coherence times exceeding several milliseconds in the presence of realistic environmental noise. Our work opens a pathway to realizing qubits with long coherence times in a new generation of quantum devices.
\end{abstract}

\section{Introduction}\label{sec:1}

The construction of a practical quantum computer inevitably requires strategies to mitigate and correct errors that arise from the inherent fragility of quantum information. Attempts to achieve fault-tolerant quantum computation have mainly focused on developing quantum error correction (QEC) protocols \cite{cai2023quantum,sivak2023real,google2025quantum}, where  information is encoded redundantly across many qubits to actively detect and correct errors. The logical error rate decreases with an increasing number of physical qubits used for encoding, but this makes the scaling of quantum computers challenging because of the qubit overhead. In addition, the successful implementation of error correction codes requires physical qubits to exhibit an error rate below the corresponding threshold.

To overcome the high resource demands from QEC, it is crucial to reduce physical error rates to enable the use of simpler and more resource-efficient error correction protocols. Consequently, parallel work has been devoted to improving the hardware itself. In this context, a primary goal is to build quantum devices that are naturally tolerant to environmental noise. Some of these efforts are spent on improving materials fabrication \cite{murray2021material, premkumar2021microscopic, chang2022reproducibility, osman2021simplified} and designs \cite{groszkowski2018coherence, gyenis2021moving}  to increase robustness to dephasing and relaxation. Alternatively, encoding quantum information non-locally provides a distinct way to decouple the qubit states from local noise, but it is often difficult to realize \cite{majumdar1969next,levitov2001quantum,callison2017protected}. 

Decoherence-Free Subspaces (DFS) offer a passive strategy to protect quantum information from environmental noise \cite{lidar1998decoherence,lidar2003decoherence,quiroz2024dynamically}, with robustness arising intrinsically from symmetry. In typical DFS constructions, this symmetry is incorporated into the couplings between the device and the environment, such that the noise induces only irrelevant global phases on the qubit states. A more powerful, yet often challenging approach is to find symmetries embedded in the system Hamiltonian itself, which offers protection regardless of the details of the environmental couplings.

In this work, we propose an interacting spin model,  that hosts a DFS spanned by the ground and first excited states, protecting in the first order from local perturbations against both relaxation and dephasing in a system of as few as four spins. In its minimal form, the spins are coupled via simple flip-flop couplings, which can be implemented in many physical platforms, such as superconducting circuits \cite{brookes2022protection,rasmussen2021superconducting}, semiconductor spins \cite{schofield2025roadmap}, neutral atoms \cite{singh2022dual, groh2016robustness, scholl2021quantum}, or trapped ions \cite{guo2024site, bruzewicz2019trapped, britton2012engineered}. We show that these states are robust against residual terms, tolerating disorder of up to 15\% in the couplings. To demonstrate a concrete pathway toward experimental realization, we map the interacting spin model to the superconducting circuit, proposing a new design of superconducting qubits and characterizing its coherence properties. 

The paper is structured as follows. First, we introduce the spin model, which provides decoherence-free states in its lowest two energy levels, and analyse the physical root of the protection. Next, we investigate the robustness of this protection when the system Hamiltonian is modified by potential residual terms or disorder that may be introduced in realistic implementations. We then map the spin model to the superconducting circuit model and identify circuit parameters required to realize the protected states. Finally, we numerically estimate the relaxation and pure dephasing times when the device is exposed to realistic environmental noises.

\begin{figure*}[th!]
    \centering
    \includegraphics[width=\linewidth]{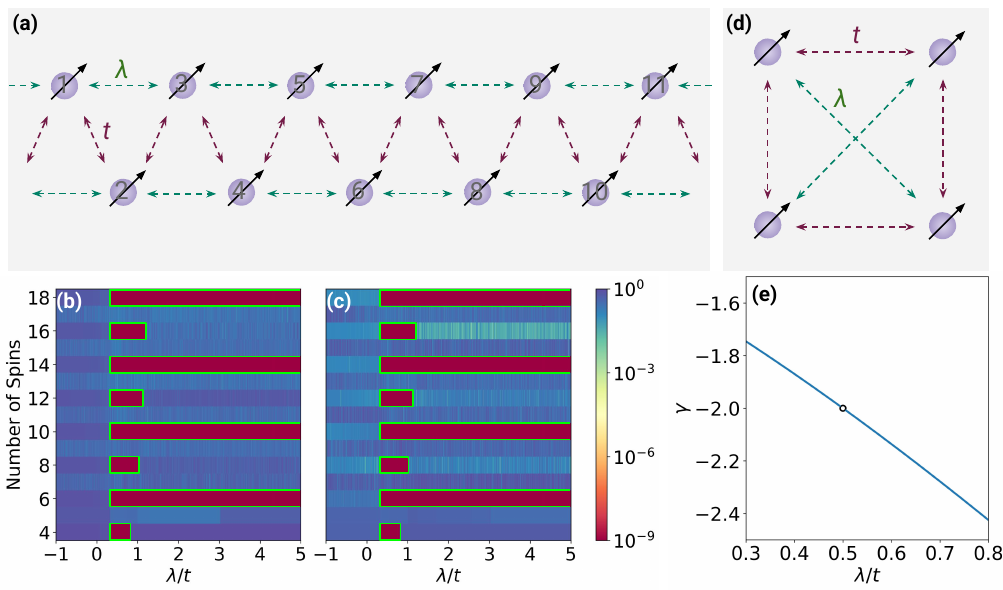}
    \caption{(a) Interacting many-body spin model with the short-range (t, red arrow) and long-range ($\lambda$, green arrow) flip-flop couplings (b) The relaxation amplitudes $\text{R} = \sqrt{\sum_{w=x,y,z} |\langle 1|\sigma_m^w|0\rangle|^2}$ versus long-range flip-flop couplings $\lambda$ for different numbers of spins. (c) The dephasing amplitudes $\text{D} = \sqrt{\sum_{w=x,y,z} |\langle 1|\sigma_m^w|1\rangle-\langle 0|\sigma_m^w|0\rangle|^2}$ as a function of $\lambda$.  In green rectangles in (b) and (c) the inversion and translation symmetries are satisfied, and the total excitation number is conserved. (d) A diagram of the finite-size spin model with a minimum of four spins. (e) The relative amplitude of the non-Bell component $\gamma$ in $|G^+\rangle$ from Eq. \ref{Eq:state1} as a function of $\lambda/t$.}
    \label{fig:1}
\end{figure*}

\section{Finite interacting spin chain model}\label{sec:2}

We propose an interacting many-body spin chain formed of $M$ qubits shown in Fig. \hyperref[fig:1]{\ref*{fig:1}(a)} desribed by the following Hamiltonian 

\begin{equation}\label{spinH}
    \begin{aligned}
            H_0 &=\frac{t}{2}\sum_{m=1}^M(\sigma_m^+\sigma_{m+1}^-+\sigma_m^-\sigma_{m+1}^+)\\
            &+\frac{\lambda}{2}\sum_{m=1}^M(\sigma_m^+\sigma_{m+2}^-+\sigma_m^-\sigma_{m+2}^+),
    \end{aligned}
\end{equation}
where $t$ and $\lambda$ are the strengths of the short-range (nearest) and long-range (next-nearest) flip-flop interactions, respectively. Periodic boundary conditions are imposed such that $\sigma_{M+1}$ = $\sigma_{1}$ and $\sigma_{M+2}$ = $\sigma_{2}$. We proceed to diagonalize the Hamiltonian to find the energy spectrum and wavefunctions. In Fig. \hyperref[fig:1]{\ref*{fig:1}(b-c)}, we plot the relaxation and pure dephasing amplitudes for the ground $|0\rangle$ and the first excited states $|1\rangle$ as a function of $\lambda/t$ for different numbers of spins ranging from 4 to 18. In specific parameter regimes, both relaxation and dephasing caused by local Pauli noises can be canceled to first order, i.e. for terms linear in $\langle i|\sigma_m^w|j\rangle$ ($w\in\{x,y,z\}$), which provides protection against local perturbations. Remarkably, this first-order cancellation still occurs with as few as four spins. While larger spin numbers can extend the DFS to a broader range of $\lambda/t$, the system also shows size modulo four effects, where for spin numbers M mod 4 = 2, this phase can be extended to $\lambda/t\rightarrow +\infty$.

\begin{figure*}[th!]
    \centering
    \includegraphics[width=0.8\linewidth]{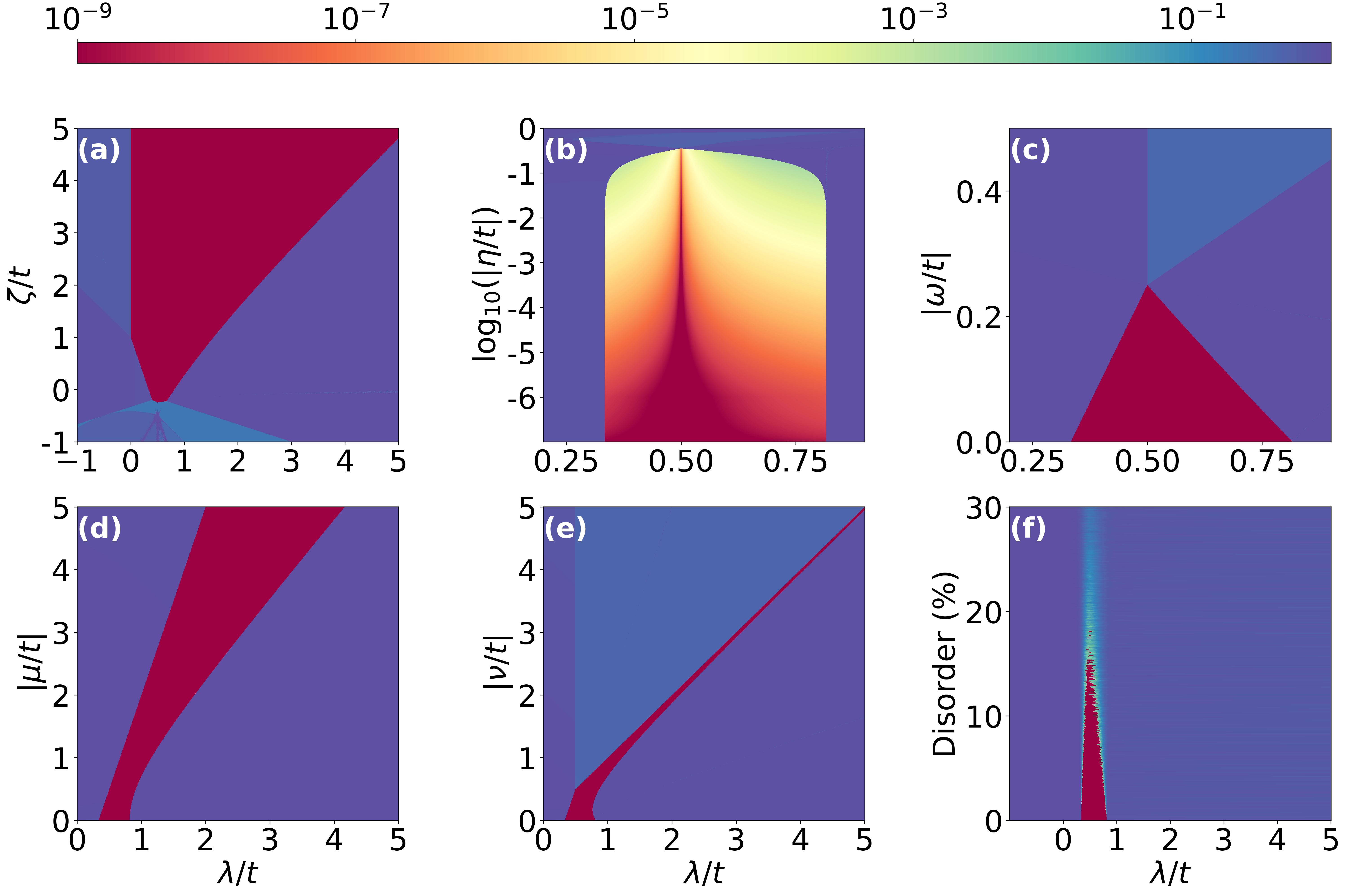}
    \caption{The combined decoherence sensitivity $\sqrt{R^2+D^2}$ as a function of long-range flip-flop coupling  $\lambda$ and (a) all-to-all 
    $\sigma^z\sigma^z$ coupling $\zeta$, (b) transverse field $\sigma^x$ amplitudes $\eta$, (c) longitudinal field $\sigma^z$ amplitudes $\omega$, (d) short-range counter-rotative terms $\mu$, (e) long-range counter-rotative terms  $\nu$, and (f) amplitudes of disorder among the couplings for $M = 4$.}
    \label{fig:2}
\end{figure*}

We consider the smallest system consisting of four spins to study the properties of the states in the decoherence-cancellation phase (Fig. \hyperref[fig:1]{\ref*{fig:1}(d)}). The wave functions of these states are crucial for understanding the origin of the disappearance of relaxation and pure dephasing. When $\lambda \neq t/2$, the ground and first excited states labeled as $|G^{\pm}\rangle$ can be analytically written as
\begin{equation}\label{Eq:state1}
    |G^+\rangle = 
        N\Big[2|\Psi^+\rangle_{1,3}|\Psi^+\rangle_{2,4} + \gamma(|\uparrow\downarrow\uparrow\downarrow\rangle + |\downarrow\uparrow\downarrow\uparrow\rangle)\Big],
\end{equation}
\begin{equation}\label{Eq:state2}
     |G^-\rangle= 
        |\Psi^-\rangle_{1,3}|\Psi^-\rangle_{2,4},
\end{equation}
where $|\Psi^{\pm}\rangle_{i,j}$ are Bell states formed between spins $i$ and $j$, the normalization factor $N = (\sqrt{4+2\gamma^2})^{-1/2}$, and $\gamma$ is the relative amplitude of the non-Bell component in $|G^+\rangle$, plotted in Fig. \hyperref[fig:1]{\ref*{fig:1}(e)} and determined by the Hamiltonian. For $\lambda<t/2$, $|G^+\rangle$ ($|G^-\rangle$) is the ground (first excited) state, and for $\lambda>t/2$, the roles are reversed. The SU(2) symmetry of the system Hamiltonian leads to the formation of multiplets. When $\lambda = t/2$, the system reaches the Majumdar–Ghosh point \cite{majumdar1969next}, where the two states become degenerate and are given by

\begin{equation}\label{Eq:state3}
    |G^{\pm}\rangle = 
    \begin{cases}
        \alpha^+ |\Psi^-\rangle_{1,3}|\Psi^-\rangle_{2,4} + 
        \beta^+ |\Psi^-\rangle_{1,2}|\Psi^-\rangle_{3,4},\\
        \alpha^- |\Psi^-\rangle_{1,4}|\Psi^-\rangle_{2,3} + 
        \beta^- |\Psi^-\rangle_{1,2}|\Psi^-\rangle_{3,4},
    \end{cases}
\end{equation}
with the values of $\alpha^{\pm}$ and $\beta^{\pm}$ given in \footnote{$\alpha^+=0.632$, $\beta^+=0.522-0.020i$, $\alpha^-=0.966$, $\beta^-=-0.062+0.014i$}. Considering $\lambda/t$ located within the decoherence-free phase, local Pauli operators can shift the states out of the computational subspace $\{|G^{+}\rangle, |G^{-}\rangle\}$ to a subspace composed exclusively of the higher energy states. This naturally leads to $\langle G^{i} |\sigma_m^{x,y,z}|G^{j}\rangle_{i,j\in\{+,-\}} = 0$, thus automatically resulting in the cancellation of relaxation and pure dephasing. We also calculate the Von Neumann entropy of the partial trace of $\rho_{\pm} = |G^{\pm}\rangle\langle G^{\pm}|$ and find $S(\rho^m_{\pm}) = 1$, where $\rho^m_{\pm}$ is the reduced state of $\rho_{\pm}$ on spin m, indicating that the spins are maximally entangled and the quantum information is stored non-locally in the system.

While Pauli noise may cause leakage from our model's protected subspace into higher-energy states when the transition frequency matches with the noise's spectral components, applying quantum optimal control \cite{chen2016measuring,werninghaus2021leakage} or state reset protocols \cite{mcewen2021removing} can suppress this type of leakage. Also, since both of the states have even parity, with a hamming weight equal to two, leakage caused by local bit flips are dectectable using parity-check measurements and can be actively corrected \cite{terhal2015quantum}.

The cancellation of relaxation and pure dephasing can be further explained by the symmetry protection. Consider the eigenstates of the number operator $\mathcal{N}$, translation operator $\mathcal{T}$, and inversion operator $\mathcal{I}$ $|0\rangle$ and $|1\rangle$, with the eigenvalues listed in Tab. \ref{Tab1}. The symmetry operators are defined by $\mathcal{N} = \sum_{m=1}^M(1+\sigma_m^z)/2$, $\mathcal{T}\sigma_m\mathcal{T}^{-1} = \sigma_{m+1}$, and $\mathcal{I}\sigma_m\mathcal{I} = \sigma_{M+1-m}$. One interesting case is where $\mathcal{T}$ has two positive unity eigenvalues while $\mathcal{I}$ has unity eigenvalues of opposite signs to each other \cite{brookes2022protection}. Here, we consider another scenario where $\mathcal{T}$ has opposite eigenvalues and $\mathcal{I}$ has identical eigenvalues. Since the condition on $\mathcal{N}$ coincides with \cite{brookes2022protection}, which leads to the cancellation of D, we only need to prove that the new conditions of $\mathcal{T}$ and $\mathcal{I}$ can lead to the cancellation of R. Let us define a unitary operator $U_i = \mathcal{T}^{2i-M-1}\mathcal{I}$ whose action is first to invert the site of spin, then translate it back to its original position. By inserting $U_i$ into the transition matrix element of any local Pauli operator $\sigma_i$, we find
\begin{equation}\label{Eq. 221}
    \begin{aligned}
        \langle 1|\sigma_i|0\rangle = \langle 1|U_i \sigma_i U_i^{-1}|0\rangle= (-1)^{-2i+M+1} \langle 1 | \sigma_i |0\rangle.
    \end{aligned}
\end{equation}
Therefore, all the transition matrix elements contributing to the relaxation are canceled: $\langle 1|\sigma_i|0\rangle = -\langle 1|\sigma_i|0\rangle = 0$, given that the system contains an even number of spins.

\begin{table}[ht]
    \centering
    \begin{tabular}{|c|c|c|}
        \hline
        Operators & $|0\rangle$ & $|1\rangle$ \\ \hline
        $\mathcal{N}$ & $M/2$ & $M/2$ \\ \hline
        $\mathcal{T}$ & $\pm 1$ & 1 \\ \hline
        $\mathcal{I}$ & $\mp 1$ & 1 \\ \hline
    \end{tabular}
    \caption{Eigenvalues of the operators $\mathcal{N}, \mathcal{T}$ and $\mathcal{I}$ for the symmetry-protected states $|0\rangle$ and $|1\rangle$. $M$ is the total number of spins in the system.}
    \label{Tab1}
\end{table}

In any realistic system, some additional residual terms may exist \cite{qiu2014coupling,pita2025blueprint,wang2017observing}. To study the effects of these residual terms, we add them to the Hamiltonian $H_0$ given in Eq. \ref{spinH}:
\begin{equation}\label{Eq:zz}
    \begin{aligned}
        H = H_0 &+ \sum_{m,n=1}^M\frac{\zeta}{4}\sigma^z_m \sigma^z_n\\
        &+ \sum_{m=1}^{M} (\eta \sigma_m^x + \omega \sigma_m^z)\\
        &+\sum_{m=1}^M \frac{\mu}{2}(\sigma_m^+\sigma_{m+1}^+ + \sigma_m^-\sigma_{m+1}^-)\\
        &+\sum_{m=1}^M \frac{\nu}{2}(\sigma_m^+\sigma_{m+2}^+ + \sigma_m^-\sigma_{m+2}^-),
    \end{aligned}
\end{equation}

\begin{figure*}[!htp]
    \centering
    \includegraphics[width=\linewidth]{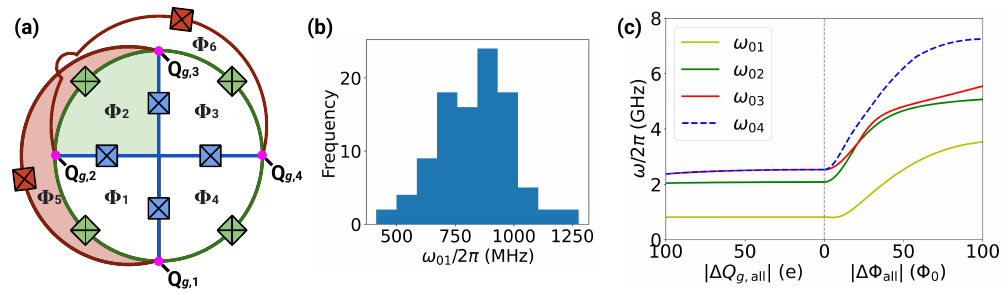}
    \caption{(a) The circuit diagram of the superconducting qubit design with four radial Josephson junctions (blue) connected by four azimuthal junctions (green) and two diametric junctions (red). Each inner loop and outer loop is threaded by a magnetic flux with amplitudes $\Phi_{i}$, and the gate charges at each island are tuned at Q$_{g,i}$. The center is grounded. (b) The distribution of the qubit transition frequency $\omega_{01}$, broadened by the disorder introduced into circuit parameters. (c) The energy spectrum of the qubit device as a function of the offset of gate charges and external magnetic flux.}
    \label{fig:3}
\end{figure*}

where $\zeta$ is the strength of the all-to-all $\sigma^z\sigma^z$ couplings, $\eta$ is the amplitude of the transverse field $\sigma^x$ and $\omega$ of the longitudinal field $\sigma^z$, $\mu$, and $\nu$ describe the strengths of the nearest and next-nearest counter-rotative terms. In Fig. \hyperref[fig:2]{\ref*{fig:2}(a)}, we calculate the combined relaxation and dephasing amplitudes versus $\lambda/t$ and $\zeta/t$ and find that the added couplings can widen the symmetry-protected phase. This occurs because the $\sigma^z\sigma^z$ couplings effectively reorder the system's energy spectrum, selectively lowering the energy of the desired higher-lying state, which with the ground state forms a symmetry-protected pair, making it the new first excited state. In Fig. \hyperref[fig:2]{\ref*{fig:2}(b)}, we find that the protection is sensitive to the amplitudes of the transverse field. The decoherence cancelation is well-preserved for $\eta/t$ being up to $10^{-1}$ when $\lambda/t \sim $ 0.5, while for other ranges of $\lambda/t$, decoherence is suppressed, though not entirely eliminated. We also find that the effect of free $\sigma^y$ is equivalent to that of $\sigma^x$. The effects of longitudinal field terms are less significant. That is, decoherence cancelation is well-preserved with $\omega/t \sim 10^{-1}$ over a large range of $\lambda/t$. The effects of counter-rotative terms are shown in Fig. \hyperref[fig:2]{\ref*{fig:2}(d,e)}, suggesting that these terms shift the protected phase, pushing the range of $\lambda/t$ to larger values as the amplitudes of these terms increase. Finally, we set all the residual terms to zero and study the effect of disorder among the coupling strengths $\lambda$ and $t$. Fig. \hyperref[fig:2]{\ref*{fig:2}(f)} shows the averaged decoherence sensitivity of many samples drawn from the modified coupling strengths $t\rightarrow t(1+\alpha)$ and $\lambda\rightarrow\lambda(1+\beta)$. Here, $\alpha$ and $\beta$ are random variables following normal distributions, with standard deviations indicated on the y axis of the plot. We find that the protection is robust to disorders up to 15 \%. Beyond this point, the state $|1\rangle$ starts to localize as we observe a decrease in the Von Neumann entropy $S(\rho_1^m)$, where $\rho_1 = |1\rangle\langle 1|$. It is also noticed that the energy gap between the $|0\rangle$ and $|1\rangle$ states increases outside the protected phase.

\section{The superconducting qubit}\label{Sec:The new Magenium qubit}

A practical design using superconducting circuits can be derived from the spin model in Eq. \ref{spinH}. As is demonstrated in Fig. \hyperref[fig:3]{\ref*{fig:3}(a)}, the spin model is mapped to the circuit model in a way that the four charge qubits composed of the radial junctions with Josephson energy $E_{Jr}$ and charging energy $E_{Cr}$ represent the four spins. The short-range and long-range flip-flop couplings between the spins are realized by the four azimuthal ($E_{Ja}, E_{Ca}$) and two diametric ($E_{Jl}, E_{Cl}$) junctions, respectively. Each small loop in the circuit has a magnetic flux bias of half flux quantum ($\Phi_{i} = \Phi_0/2$), controlling the type of interaction between qubits. The intrinsic capacitances of the Josephson junctions will contribute $\sigma^z\sigma^z$ couplings \cite{brookes2022protection,hita2022ultrastrong}. By tuning the gate charge of each island (except the grounded central island) to half-cooper pair ($Q_g = 1e$) and forcing $E_{Ca} \ll E_{Cr}$, the $\sigma^z\sigma^z$ couplings can be approximated as all-to-all with the same amplitude. We identify the above conditions (flux and gate charge) as the optimal point for the qubit. Details of the derivation of the circuit Hamiltonian are provided in Appx. \hyperref[appendixA]{A}.

\begin{figure*}[btp]
    \centering
    \includegraphics[width=0.8\linewidth]{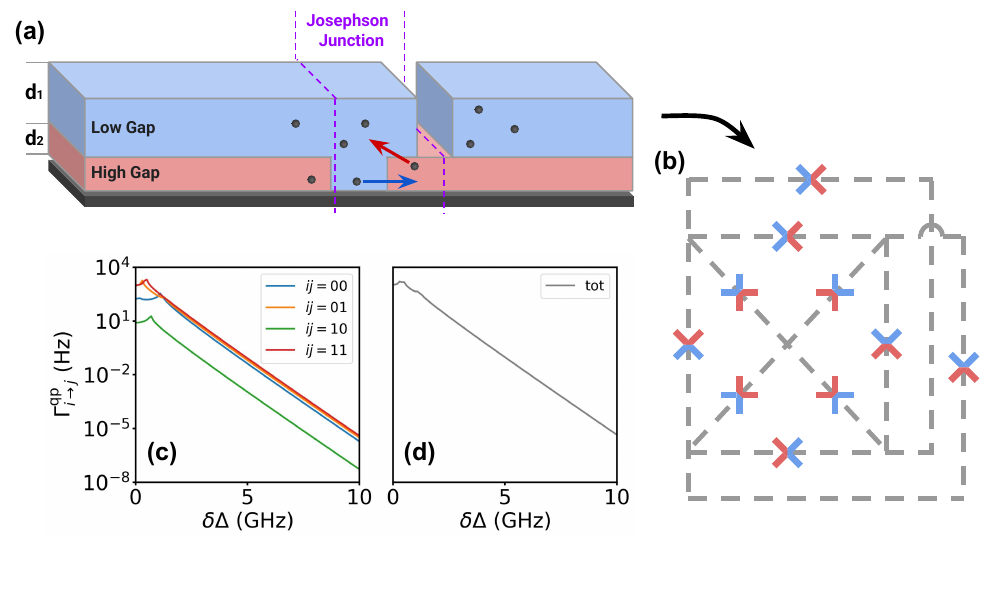}
    \caption{(a) A schematic diagram of the Josephson junction structure formed by two layers of aluminum films with different thicknesses. The pink layer is thinner and has a higher superconducting gap. The quasiparticle can tunnel from both sides. (b) The direction of the gap engineering used for the Josephson junctions in our design. (c) The parity switching rates versus the gap difference $\delta \Delta$, while fixing the average superconducting gap of two layers at $\bar{\Delta}$ = 50 GHz, with the index i indicating the initial state of the qubit at the operation point, and the index j represents the final state after the quasiparticle tunneling. (d) The total rate $\Gamma^{\text{qp}}_{\text{tot}} = 0.5(\Gamma^{\text{qp}}_{0\rightarrow0}+\Gamma^{\text{qp}}_{0\rightarrow1}+\Gamma^{\text{qp}}_{1\rightarrow0}+\Gamma^{\text{qp}}_{1\rightarrow1})$ versus the gap difference $\delta \Delta$.}
    \label{fig_4}
\end{figure*}

We choose the following parameters to demonstrate the properties of the qubit: $E_{Jr}$ = 2.5 GHz, $E_{Cr}$ = 5 GHz, $E_{Ja}/h = 5$ GHz, $E_{Ca}/h = 2.5$ GHz, $E_{Jl}/h = 5.55$ GHz, and $E_{Cl}/h = 2.25$ GHz, which gives the transition frequency $\omega_{01}/2\pi$ = 818 MHz. This set of parameters is located within the symmetry-protected phase (details of sweeping the parameters are given in Appx. \hyperref[appendixB]{B}). In realistic devices, the disorder among the circuit parameters is inevitable. We generate samples of circuit parameter sets, each defined by the four normally distributed random variables $\alpha, \beta, \gamma$, and $\delta$, with means of zero and standard deviations of 2\% for $\alpha$ and $\beta$, 0.2\% for $\gamma$, and 0.1\% for $\delta$. For each sample, the Josephson energies of the junctions become $E_J(1+\alpha)(1+\beta)$, and the corresponding charging energies are replaced by $E_C/(1+\alpha)$. We substitute the magnetic flux bias in each loop with $\Phi_{i}(1+\gamma)$ and gate charges with $N_g(1+\delta)$. The transition frequency is widened by the disorder and is shifted by 18 MHz (Fig. \hyperref[fig:3]{\ref*{fig:3}(b)}). The effects of each type of disorder on the transition frequency are given in Appx. \hyperref[appendixE]{E}.

We plot the transition frequencies of the first four excited states of the qubit device ($\omega_{0j}$ for $j \in \{1,2,3,4\}$) versus the offsets of gate charges and magnetic flux biases in Fig. \hyperref[fig:3]{\ref*{fig:3}(c)}. While they are insensitive to the shifts of gate charges, the transition frequencies increase as the flux moves away from the optimal point. We also show the case where both gate charges and magnetic fluxes are shifted away from the optimal point at the same time in Appx. \hyperref[appendixB]{B}. 

\section{Coherence properties}\label{Sec:Coherence properties}

The relaxation of the qubit can be attributed to the dielectric losses. We consider the dielectric loss from both the intrinsic capacitance of the Josephson junctions and the geometric capacitance between the metallic islands (Appx. \hyperref[appendixD]{D}). For the intrinsic capacitance of Josephson junctions, we choose the loss tangent $\tan\delta_{\text{J}}=10^{-7}$. For the geometric capacitance, assuming using sapphire substrates, we use $\tan\delta_{\text{geo}}=10^{-6}$ \cite{kusunoki2002dielectric,read2023precision,deng2023titanium}. At the optimal point, the relaxation rate from dielectric loss is zero, as it is proportional to the transition matrix element $\langle 1|\hat{n}|0\rangle$, which vanishes due to our design. However, the unavoidable variations between the circuit parameters can break the circuit symmetries and result in non-zero $\langle 1|\hat{n}|0\rangle$.

\begin{figure}
    \centering \includegraphics[width=\linewidth]{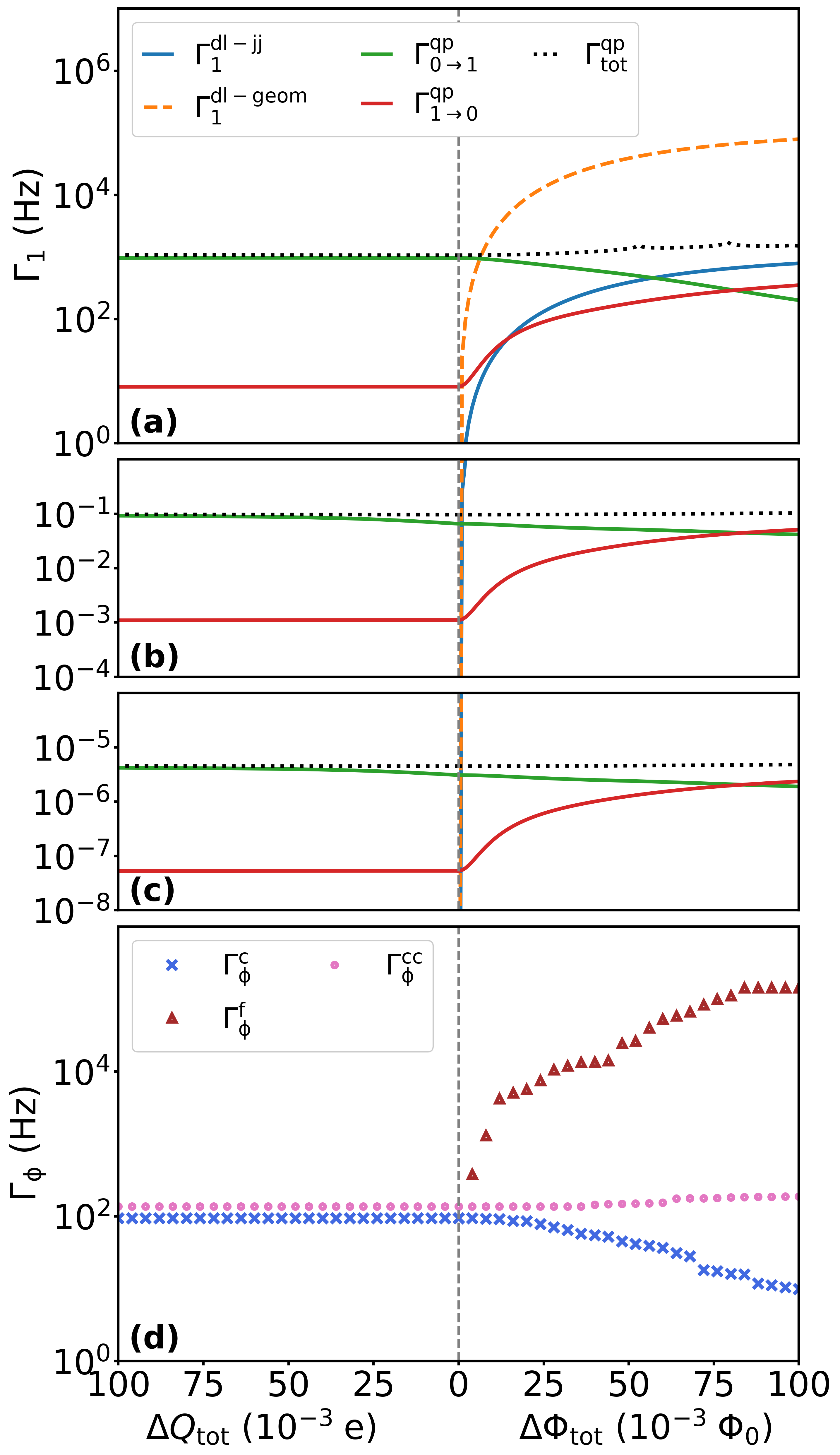}
    \caption{(a-c) Qubit relaxation and excitation rates contributed by dielectric losses from Josephson junctions $(\Gamma^{\text{dl-jj}}_1)$, geometric capacitance $(\Gamma^{\text{dl-geom}}_1)$ and quasiparticle tunnelings $(\Gamma^{\text{qp}})$ as a function of the total offsets of gate charges and magnetic flux biases. Several gap differences are used: (a) $\delta\Delta = 0$; (b) $\delta\Delta = 5$ GHz; (c) $\delta\Delta = 10$ GHz. (d) Pure dephasing rates contributed by charge noise ($\Gamma_{\upphi}^{\text{c}}$), flux noise ($\Gamma_{\upphi}^{\text{f}}$) and critical current noise ($\Gamma_{\upphi}^{\text{cc}}$) on total offsets in gate charges and magnetic flux biases. The lower boundary is 1 Hz, and data smaller than 1 Hz is not shown in this figure. }
    \label{fig_5}
\end{figure}

Quasiparticle tunnelings can lead to parity-switching of the gate charges, along with possible state relaxation. For example, an electron tunneling across one of the azimuthal junctions will modify the gate charges from $(0.5,0.5,0.5,0.5)$ to $(0.0,1.0,0.5,0.5)$ and shifts the qubit transition frequency. To avoid parity changes, we propose to fabricate the qubit to have various layers of different widths through gap engineering, which can suppress the tunneling rates by several orders of magnitude \cite{kamenov2023suppression}. The circuit can be constructed with two layers of aluminum films with different thicknesses, leading to different superconducting gaps. The two sides of the Josephson junctions are made from these two different materials with a gap difference $\delta \Delta$ (Fig. \hyperref[fig_4]{\ref*{fig_4}(a)}). In Fig. \hyperref[fig_4]{\ref*{fig_4}(b)}, we apply this structure to the Josephson junctions in our qubit, showing their arrangement in the device, with different colors indicating the direction of the gap engineering. Using the theoretical framework of \cite{diamond2022distinguishing}, we estimate the tunneling rates and demonstrate their suppression through gap engineering (details given in Appx. \hyperref[appendixC]{C}). In Fig. \hyperref[fig_4]{\ref*{fig_4}(c,d)}, we plot the tunneling rates as a function of the gap difference for four different scenarios. This reveals that quasiparticle tunneling not only relaxes the qubit state but has an even higher chance of exciting it ($\Gamma_{0\rightarrow1}^{
\text{qp}
} = 938 \text{ Hz} \gg \Gamma_{1\rightarrow0}^{
\text{qp}
} = 8 \text{ Hz}$). Nevertheless, we find that the tunneling rates can be suppressed by orders of magnitude with a gap difference of around several GHz. With $\delta\Delta$ = 10 GHz, the excitation rate is reduced below $10^{-5}$ Hz, the relaxation rate is reduced below $10^{-7}$ Hz, and the total tunneling rate is suppressed to $10^{-5}$ Hz, equivalent to $\sim$ 1.2 days. With the suppressed tunneling rates, to recalibrate the gate voltages to maintain the qubit at the optimal point, we can actively monitor the parity change of the qubit. Recent developments in shielding and filtering techniques suppress the generation of photon-induced quasiparticles \cite{pan2022engineering,gordon2022environmental,diamond2024quasiparticles}. 

There are three channels that contribute to the dephasing rate of the qubit: charge noise, flux noise, and critical current noise. The noise sources are assumed to have a 1/f spectrum with upper and lower cutoffs equal to 1 MHz and 1 Hz, respectively. For charge noise and flux noise, we use the total strengths $A_{\text{c}} = (2 \times 10^{-4} \;e)^2$ and $A_{\text{f}} = (2 \times 10^{-6}\; \Phi_0)^2$, which are distributed evenly among the gate charges and circuit loops, respectively. We find the pure dephasing time $T_{\upphi}^{\text{c}} = 10.5$ ms for charge noise and $T_{\upphi}^{\text{f}} \approx 7.4$ s for flux noise. We choose $A_{\text{cc}} = 1\times10^{-7} I_{\text{c}}$ for critical current noise \cite{groszkowski2018coherence, chang2023tunable}, where $I_{\text{c}}$ is the critical current of each Josephson junction, then find $T_{\upphi}^{\text{cc}} = 7.3 $ ms. In Fig. \ref{fig_5}, we find that small offsets in gate charges have negligible effects on the coherence properties of the qubit. The dielectric loss from Josephson junctions and geometric capacitances, the relaxation rates from quasiparticle tunnelings, and pure dephasing rates from flux noises all increase with the offset of flux bias. Conversely, the excitation rates from quasiparticle tunneling and pure dephasing caused by charge noises decrease when the flux bias offsets increase. The methods used to calculate the relaxation and dephasing rates are explained in Appx. \hyperref[appendixD]{D}.

The qubit's coherence times estimated by us assume that disorder in the Josephson junctions, circuit loop areas, and gate charges are all negligible. However, disorder will inevitably be introduced during the fabrication process and due to the drifting charges in the substrate. Applying the same approach in the previous section, the effects of each type of disorder are calculated and summarized in Appx. \hyperref[appendixE]{E}. With all the disorders being applied to the samples together, we find $T_{0\rightarrow 1}^{\text{qp}}$ unaffected and $T_{1\rightarrow 0}^{\text{qp}}$ reduced to 23.7 ms. However, both are kept above 10 s for $\delta\Delta$ = 5 GHz and above $10^5$ s for $\delta\Delta$ = 10 GHz. The relaxation rate from dielectric loss is no longer zero, and the corresponding relaxation time is reduced to $T_1^{\text{dl}}$ = 15.2 ms. The dephasing time of critical current noise stays unaffected while the other two are decreased due to the disorder but are still at the millisecond range: $T_{\upphi}^{\text{c}}=3.4$ ms, $T_{\upphi}^{\text{f}}=9.2$ ms.


\section{Conclusions}
In conclusion, we have presented a practical blueprint for an intrinsically protected qubit, implemented using the lowest two energy states of a simple interacting spin model. It is remarkable that symmetry protection can be achieved with a minimal system of just four spins, and this protection is robust enough to tolerate up to 15\% of disorder among the couplings. This resource efficiency offers an alternative to traditional QEC protocols, which generally demand a large overhead of physical qubits. This advantage makes the passive protection of quantum devices a near-term strategy and enables the scaling up of practical quantum computers.

We demonstrated a physical realization of the protected model using superconducting circuits. However, the simplicity of the design makes it suitable for a wider range of physical platforms. Moving forward requires developing high-fidelity single- and two-qubit gates within the DFS and implementing the model in other promising architectures. As a first step, we propose a controlling scheme using STIRAP pulses to drive the transitions of the qubit via higher energy states of the device  (Appendix \hyperref[appendixE]{E}). Ultimately, our work provides a foundational building block for a new generation of intrinsically noise-protected quantum hardware, offering a more resource-efficient strategy toward fault-tolerant quantum computation.

\section*{Acknowledgements}
This work was supported by the Engineering and Physical Sciences Research Council [grant number EP/S021582/1]. E.G. acknowledges support from EPSRC grant EP/T001062/1. The authors acknowledge support from the QuantERA project PROTEQT. The authors acknowledge the use of the UCL Myriad High Performance Computing Facility (Myriad@UCL), and associated support services, in the completion of this work.

Y.S., E.G., M.S., and M.S. declare a relevant patent application: United Kingdom Patent Application No. 2515373.5.

\appendix

\section{Circuit Hamiltonian}\label{appendixA}

To obtain the circuit Hamiltonian, we first construct the Lagrangian $L = T- V$, consisting of the kinetic energy $T$, describing the capacitive charging energies, and the potential energy $V$ representing the inductive energies stored in the Josephson Junctions. We write the node flux as $\phi_i$ for site i. The kinetic part can be obtained by summing up all the capacitive charging energies:
\begin{equation}
    \begin{aligned}
        T &= \frac{1}{2} \sum_{i=1}^4 [C_r \dot{\phi}_{i}^2 + C_a (\dot{\phi}_{i+1}-\dot{\phi}_{i})^2 + \frac{1}{2} C_l (\dot{\phi}_{i+2}-\dot{\phi}_{i})^2]\\
        &= \frac{1}{2} \sum_{i,j=1}^4 C_{i,j}\dot{\phi}_i\dot{\phi}_j,
    \end{aligned}
\end{equation}
where $C_{i,j}$ are the matrix elements of the capacitance matrix given by
\begin{widetext}
\begin{equation}\label{C_matrix_Ccouple_multisite}
    \textbf{C} = \left(
    \setlength{\arraycolsep}{2pt} 
    \begin{array}{cccc}
        C_r + 2C_a + C_l & -C_a & -C_l & -C_a \\
        -C_a & C_r + 2C_a + C_l & -C_a & -C_l \\
        -C_l & -C_a & C_r + 2C_a + C_l & -C_a \\
        -C_a & -C_l & -C_a & C_r + 2C_a + C_l 
    \end{array}
    \right).
\end{equation}
\end{widetext}

The potential part includes the inductive energy terms of all the Josephson Junctions. The fluxoid quantization condition requires that the flux difference across all the components on the loop plus the external flux threading the loop equal an integer numbers of magnetic flux quanta \cite{krantz2019quantum}. The flux difference across the azimuthal $\Delta \phi_{a,i}$ and diametric junctions $\Delta \phi_{l,i}$ is
\begin{equation}
    \begin{aligned}
        \Delta \phi_{a,i} &= \phi_{i+1} - \phi_i - \Phi_{int},\\
        \Delta \phi_{l,i} &= \phi_{i+2} - \phi_i - 2\Phi_{int} - \Phi_{ext},
    \end{aligned}
\end{equation}
and the potential reads
\begin{equation}
    \begin{aligned}
        V &= -E_{Jr}\sum_{i=1}^4\cos(\frac{\phi_i}{\varphi_0})-E_{Ja}\sum_{i=1}^4\cos(\frac{\phi_{i+1}-\phi_i-\Phi_{int}}{\varphi_0}) \\&-E_{Jl}\sum_{i=1}^2\cos(\frac{\phi_{i+2}-\phi_i-2\Phi_{int}-\Phi_{ext}}{\varphi_0}).
    \end{aligned}
\end{equation}
In the Lagrangian formalism, the circuit Hamiltonian is described by the variables $\phi_i$, $\dot{\phi}_i$. The conjugate momentum variable $Q_i$ of $\phi_i$ is attainable from the partial derivatives of the Lagrangian:
\begin{equation}\label{variable_Q}
    Q_i = \frac{\partial L}{\partial \dot{\phi_i}} = \sum_{j=1}^4 C_{ij}\dot{\phi_j}.
\end{equation}
Using the conjugate momentum $Q_i$, we apply the Legendre transformation and write the circuit Hamiltonian as
\begin{equation}
    \begin{aligned}
        H &= \sum_{i=1}^4 \dot{\phi}_i Q_i - L = \frac{1}{2}\sum_{i,j=1}^4 C_{ij}^{-1} Q_i Q_j + V\\
        & = T + V.
    \end{aligned}
\end{equation}
We can quantize the Hamiltonian by replacing $Q_i$ with $2e(\hat{N}_i - N_{g,i})$ and $\phi_i/\varphi_0$ with $\hat{\theta}_i$, where $\hat{\theta}_i$ and $\hat{N}_i$ satisfy the commutation relation $[\hat{\theta}_i,\hat{N}_j] = i\hbar$. $\hat{N}_i$ is the charge number operator, $\theta_i$ is the phase operator, $N_{g,i}$ is the gate charge at site i, and $C_{ij}^{-1}$ are the matrix entries of the inverse of the capacitance matrix defined in Eq. \ref{C_matrix_Ccouple_multisite}. The total Hamiltonian can be expressed as
 \begin{equation}
    \begin{aligned}
            H &= 2e^2 \sum_{i,j=1}^4 C_{ij}^{-1} (\hat{N}_i-N_{g,i}) (\hat{N}_j-N_{g,j})\\
        &-E_{Jr}\sum_{i=1}^4\cos(\hat{\theta}_i) -E_{Ja}\sum_{i=1}^4\cos(\hat{\theta}_{i+1}-\hat{\theta}_i-\Phi_{int}/\varphi_0)\\
        &-E_{Jl}\sum_{i=1}^2\cos(\hat{\theta}_{i+2}-\hat{\theta}_i-(2\Phi_{int}+\Phi_{ext})/\varphi_0).
    \end{aligned} 
 \end{equation}
For the kinetic part, we take the limit $C_r/C_a \rightarrow 0$, such that all of the elements tend toward a constant $C_{ij}^{-1}\rightarrow \frac{1}{4C_r}$, allowing all the kinetic terms to share the same amplitude. By fixing all the gate charges $N_{g,i}$ to half integers and reducing the system to two charge states per node, we can identify $\hat{N}_i - N_{g,i}$ with $\frac{1}{2}\sigma^z$, resulting in couplings that take the same form as the all-to-all $\sigma_i^z\sigma_j^z$ couplings.

\begin{figure*}[htp!]
    \centering
    \includegraphics[width=0.6\linewidth]{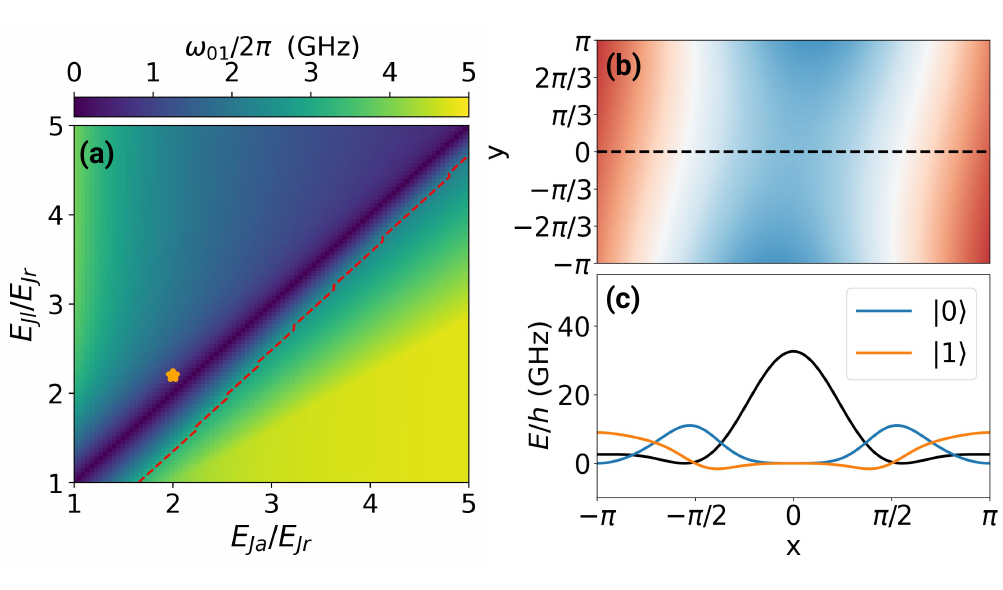}
    \caption{(a) The transition frequency between the lowest states $|0\rangle$ and $|1\rangle$ as a function of $E_{Jl}/E_{Jr}$ and $E_{Ja}/E_{Jr}$. The red dashed line separates the symmetry-protected (upper left) and unprotected (lower right) phases. Degeneracy between the two states is  when $E_{Jl} = E_{Ja}$. The case marked by the orange star is discussed in the main text. (b) The potential energy of the circuit in the (x,y) plane with global minima localized at the two deformed white valleys. (c) Potential energy (black curve) for y = 0 as a function of x and the projection of the ground state wave functions.}
    \label{fig_b1}
\end{figure*}

 In the charge number basis, The phase variables in the potential part can be expressed in terms of the tunneling and anti-tunneling operators
\begin{equation}
    \cos(\hat{\theta}_i) = \frac{1}{2}(\Sigma_i^++\Sigma_i^-),
\end{equation}
\begin{equation}
    \sin(\hat{\theta}_i) = \frac{1}{2i}(\Sigma_i^+-\Sigma_i^-),
\end{equation}
where the tunneling and anti-tunneling operators are given by
\begin{equation}
    \Sigma_i^+ = \sum_n |n+1\rangle_i\langle n|_i,
\end{equation}
    \begin{equation}
    \Sigma_i^- = \sum_n |n\rangle_i\langle n+1|_i.
\end{equation}
If we restrict each node to two charge states, we can equate the $E_{Jr}$ terms in the circuit Hamiltonian to the transverse field terms $\sigma_x$ in the spin Hamiltonian. Similarly, if we tune the magnetic flux in each loop to half flux quantum ($\Phi_{int}/\varphi_0 = \Phi_{ext}/ \varphi_0 = \pi$), the flip-flop couplings in the spin model are the same as the $E_{J_a}$ and $E_{J_l}$ terms once expanded

\begin{equation}
    \cos(\hat{\theta}_{i+1}-\hat{\theta}_{i}-\Phi_{int}/\varphi_0)
    =-\frac{1}{2} (\Sigma_i^+\Sigma_{i+1}^-  + \Sigma_i^-\Sigma_{i+1}^+),
\end{equation}

\begin{equation}
    \cos(\hat{\theta}_{i+2}-\hat{\theta}_{i}-(2\Phi_{int}+\Phi_{ext})/\varphi_0)
    =-\frac{1}{2} (\Sigma_i^+\Sigma_{i+2}^-  + \Sigma_i^-\Sigma_{i+2}^+ ).
\end{equation}

\section{Parameters of the qubit and qubit states}\label{appendixB}

\begin{figure}[h!]
    \centering
    \includegraphics[width=0.8\linewidth]{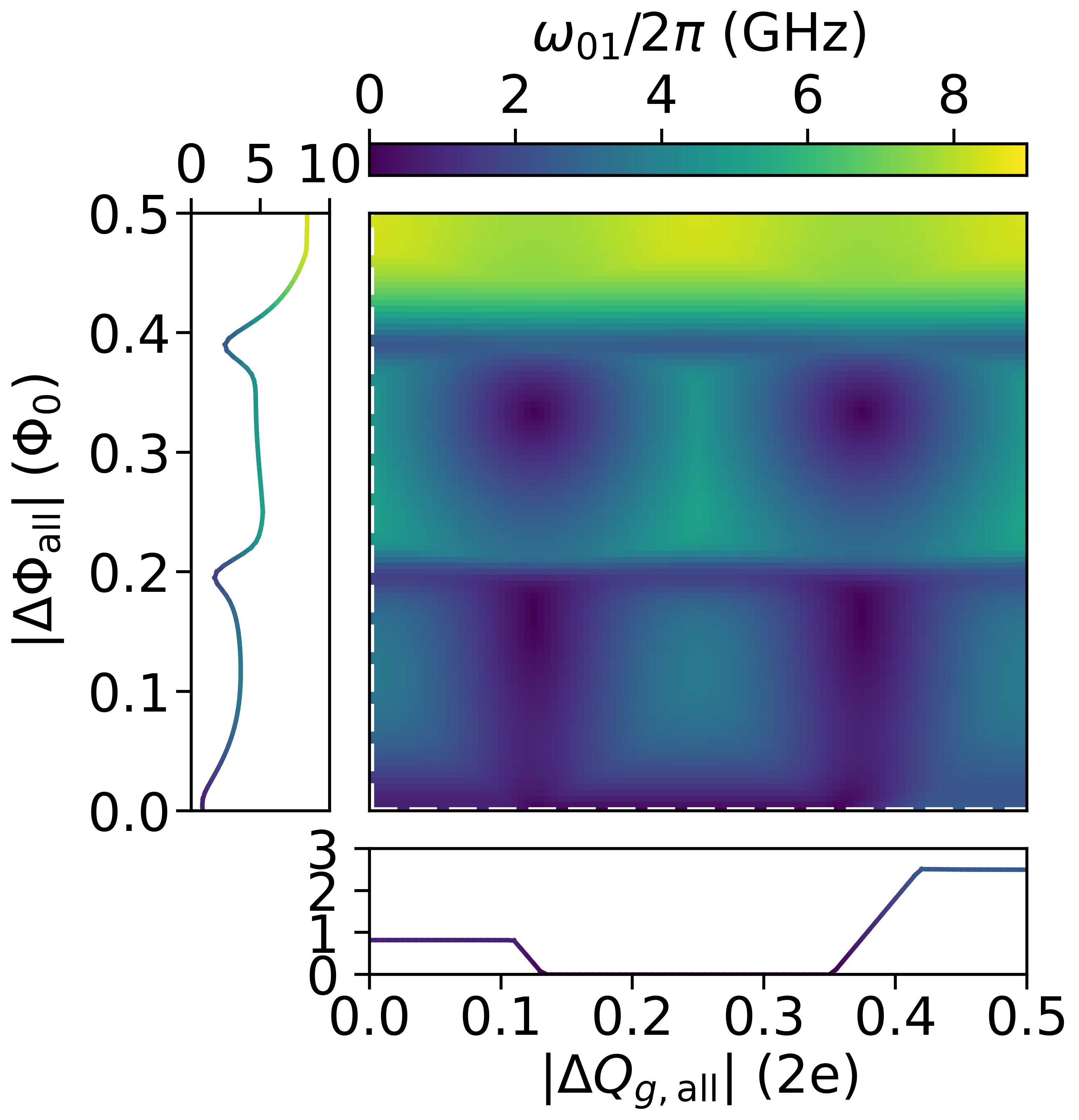}
    \caption{Transition frequency $\omega_{01}$ of the qubit as a function of the gate charge and magnetic flux bias offsets. All the gate charges and flux biases vary simultaneously.}
    \label{fig_b2}
\end{figure}

For the Josephson junctions in the qubit, we choose the plasma frequency $\sqrt{8 E_J E_C}$ = 10 GHz and fix the parameters of the radial junctions to $E_{Jr}$ = 2.5 GHz while varying the parameters of the other two junctions $E_{Ja}$ and $E_{Jl}$ to search for the symmetry-protected ground states in the circuit model. The ground states are obtained by the exact diagonalizations of the circuit Hamiltonian using the Python software PRIMME \cite{PRIMME,svds_software}. The symmetry-protected phase and transition frequency $\omega_{01}$ are plotted in Fig. \hyperref[fig_b1]{\ref*{fig_b1}(a)}. To find the operatable parameters for the qubit, we have to select the circuit parameters that lead to $\omega_{01}, \omega_{12} > k_B T$, to suppress the thermal excitations, where $T$ is the device temperature. To understand these states in the circuit model, we write the node phases $\theta_n$ in terms of 2 variables $x$ and $y$: $\theta_n = nx + y$. The potential of the qubit as a function of $x$ and $y$ is plotted in Fig. \hyperref[fig_b1]{\ref*{fig_b1}(b)}. Two deformed valleys of global minima are observed close to $x  = \pm \pi/2$. We plot the potential for $y = 0$ in Fig. \hyperref[fig_b1]{\ref*{fig_b1}(c)} along with the projected wave functions of the two ground states. The ground state $|0\rangle$ are symmetric superpositions of clockwise and anti-clockwise persistent current states localized at the global potential minima $\pm 0.55\pi$ while the $|1\rangle$ state is centralized at $\pm\pi$. In Fig. \hyperref[fig_b2]{\ref*{fig_b2}}, we plot the dependency of the qubit transition frequency $\omega_{01}$ on both the gate charges and magnetic flux biases.

\section{Gap engineering details}\label{appendixC}

Here, we explain the details of the model used to calculate the effects of gap engineering on our qubit design. The superconducting gap has an inverse dependence on the thickness of the aluminum film and is estimated by $\Delta/h \approx 43.5 \text{ GHz} + \frac{145 \text{ GHz} \cdot\text{nm}}{d}$, where $d$ is the thickness of the layer. Quasiparticle tunneling can lead to the parity switching of gate charges along with possible qubit state transitions, e.g. excitation from ground state $|0\rangle$ to $|1\rangle$, or relaxation from $|1\rangle$ to $|0\rangle$. The quasiparticles may tunnel without exchanging energy with the qubit. In this case, the qubit states remain unchanged. The tunneling rates for each type of occurrence are given by
\begin{equation}\label{Eq_qp3}
    \Gamma^{ij} = \Gamma' (|\langle j|\cos \frac{\hat{\varphi}}{2}|i\rangle|^2 S_{-}^{ij} + |\langle j|\sin \frac{\hat{\varphi}}{2}|i\rangle|^2 S_{+}^{ij}),
\end{equation}
where $\Gamma' = \frac{16 E_J}{\pi\hbar}$ is the tunneling amplitude, and $S^{ij}_{\pm}$ are the structure factors that decrease exponentially with the gap difference $\delta\Delta$, and also depend on the transition frequency $\omega_{ij}$, the device temperature $T$, and the quasiparticle density $x_{qp}$ \cite{diamond2024quasiparticles}. The quasiparticles could tunnel through any Josephson junctions from both sides. We assume the quasiparticle density at each side of every junction to be $x_{\text{qp}} = 5 \times 10^{-9}$ \cite{brookes2021protected, diamond2022distinguishing, diamond2024quasiparticles} and count all possible tunneling events, assuming the initial preparation of the qubit at its operation point.

\section{Coherence time calculation}\label{appendixD}

The dielectric loss in the qubit comes from the intrinsic capacitance of Josephson junctions, and the geometric capacitance originating from the semiconductor substrate. The dielectric-loss-induced relaxation rate can be written as
\begin{equation}
    \Gamma_{1}^{\text{dl}} = \frac{2}{\hbar} \tan \delta \;\text{Tr} (\textbf{C}^{-1}\textbf{Q}^2),
\end{equation}
where $\tan \delta$ is the loss tangent determined by the material, $\textbf{C}$ is the capacitance matrix. For the intrinsic capacitance, $\textbf{C}$ is given in Eq. \ref{C_matrix_Ccouple_multisite}. $\textbf{Q}^\textbf{2}$ is a matrix composed of the quantum overlaps of the charge operators \cite{stern2014flux}. The matrix entries are given by
\begin{equation}
    \textbf{Q}^\textbf{2}_{i,j} = \langle 1|\hat{Q}_i|0\rangle\langle 0|\hat{Q}_j |1\rangle.  
\end{equation}

\begin{figure}[h!]
    \centering
    \includegraphics[width=0.5\linewidth]{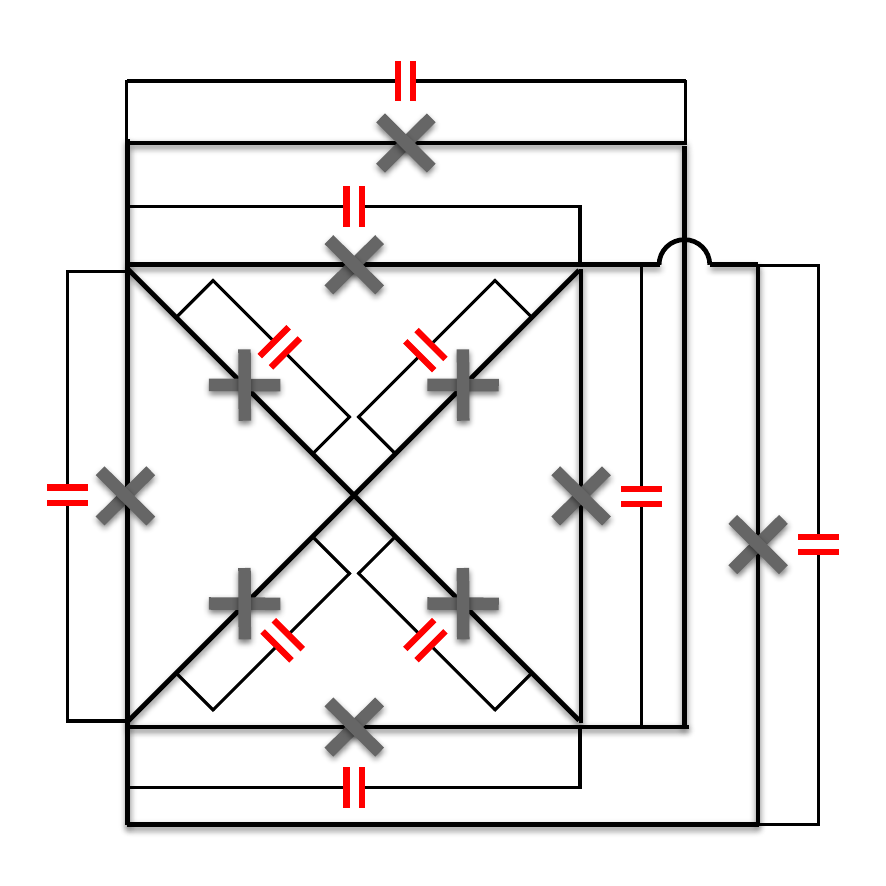}
    \caption{Circuit diagram including the geometric capacitances (red).}
    \label{fig_d1}
\end{figure}

To estimate the effects of the geometric capacitance on the dielectric loss, we add extra capacitances parallel to the Josephson junctions, each equal to 10\% of the capacitance of the parallel junctions (Fig. \ref{fig_d1}). It is checked that the symmetry protection is unaffected by these extra capacitances. 

The pure dephasing time is estimated based on the assumption that the noise has a lower frequency than the qubit transition frequency, thus is unable to cause state transitions. The noise can then be considered a parameter in the qubit Hamiltonian that changes adiabatically. We assume that at moment t, the transition frequency between $|0\rangle$ and $|1\rangle$ is changed by the noise by $\delta \omega_{01}[\vec{x}(t)]$, where $\vec{x}(t)$ is the noise written in the form of a vector. When the qubit is exposed to the noise, the phase difference $\phi_{01}(t) = \int_0^t \delta\omega_{01}[\vec{x}(t')]dt'$ fluctuates between the qubit state $|0\rangle$ and $|1\rangle$, resulting in destructive interference among these variations, leading to the dephasing of the qubit \cite{ithier2005manipulation}. The dephasing time can be estimated by finding the time it takes for the decay function $f(t)  = \langle \exp[-i\phi_{01}(t)]\rangle$ to drop from 1 to 1/e.

To calculate the frequency variation $\delta\omega_{01}(t)$ at every time t is too computationally expensive. We take many samples of the noise vectors $\vec{x}(t)$ and calculate the corresponding $\delta\omega_{01}$, then use Taylor expansion to construct the relation between $\delta\omega_{01}(t)$ and $\vec{x}(t)$ expressed by
\begin{equation}
    \delta\omega_{01}(t) = \textbf{g}\vec{x}(t) + \frac{1}{2} \vec{x}(t)^T\textbf{h}\vec{x}(t) + \mathcal{O}[\vec{x}(t)^3],
\end{equation}
where $\textbf{g}$ is the gradient vector and $\textbf{h}$ is the Hessian matrix. Examples of noise can be obtained by sampling elements following the Power Spectral Density (PSD) in the frequency domain and applying the inverse Fourier transform to construct the noise in the time domain. The PSD is divided into N bins with width $\Delta f = 0.5 \text{ Hz}$. The corresponding amplitudes are determined by
\begin{equation}
    X_{m,k} = Z_{m,k}\sqrt{S_{x_m} (f_k)\Delta f},
\end{equation}
where $f_k = k\Delta f$ and $Z_{m,k}$ is sampled from complex normal distribution
\begin{equation}
    Z_{m,k} \sim \frac{1}{\sqrt{2}}[\mathcal{N}(0,1)+i\mathcal{N}(0,1)].
\end{equation}
The signal can then be constructed from
\begin{equation}
    x_m(t_l) = \sum_{k=\frac{1-N}{2}}^{k=\frac{N-1}{2}} X_{m,k}\exp(i2\pi f_kt),
\end{equation}
where $t_l = l\Delta t $ and the time step $\Delta t = 1/N\Delta f$. N has to be odd (N = $2\times10^6-1$) to impose $Z_{m,k} = Z^*_{m,-k}$, making the signal real. For the PSD, we use $S_{x_m} = A$ for $|f|< 1\text{ Hz}$ and $S_{x_m} = 0$ for $|f| > 1 \text{ MHz}$. For $1 \text{ Hz} < |f| < 1 \text{ MHz}$, the PSD is given by the 1/f noise where $S_{x_m} = A/|f|$, where A is the total noise strengths. 

\section{Effects of disorder on the coherence time}\label{appendixE}

\begin{figure*}
    \centering
    \includegraphics[width=0.7\linewidth]{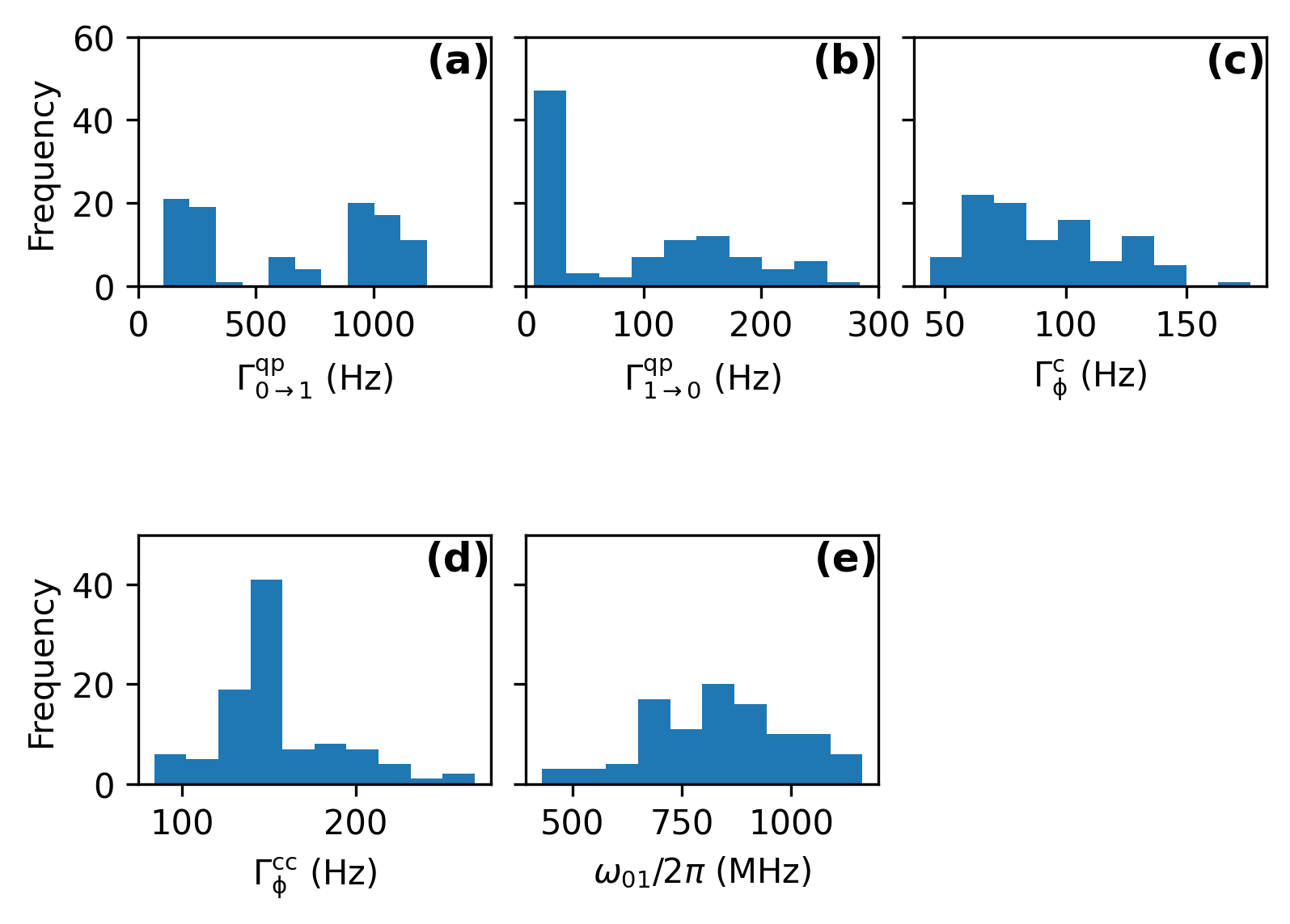}
    \caption{Histograms (with 2\% variation applied to the junction parameters) of (a) excitation (b)  relaxation rates from quasiparticle tunnelings $T_{0\rightarrow1 }^{\text{qp}} =665$ Hz and $T_{1\rightarrow 0}^{\text{qp}} =90$ Hz, (c) pure dephasing rate contributed by charge noise with mean of 90 Hz, (d) pure dephasing rate from critical-current noise with mean of  153 Hz. All the other decoherence rates are not affected. (e) Transition frequency $\omega_{01}$ with mean of 835 MHz.}
    \label{fig_e1}
\end{figure*}

\begin{figure*}
    \centering
    \includegraphics[width=0.7\linewidth]{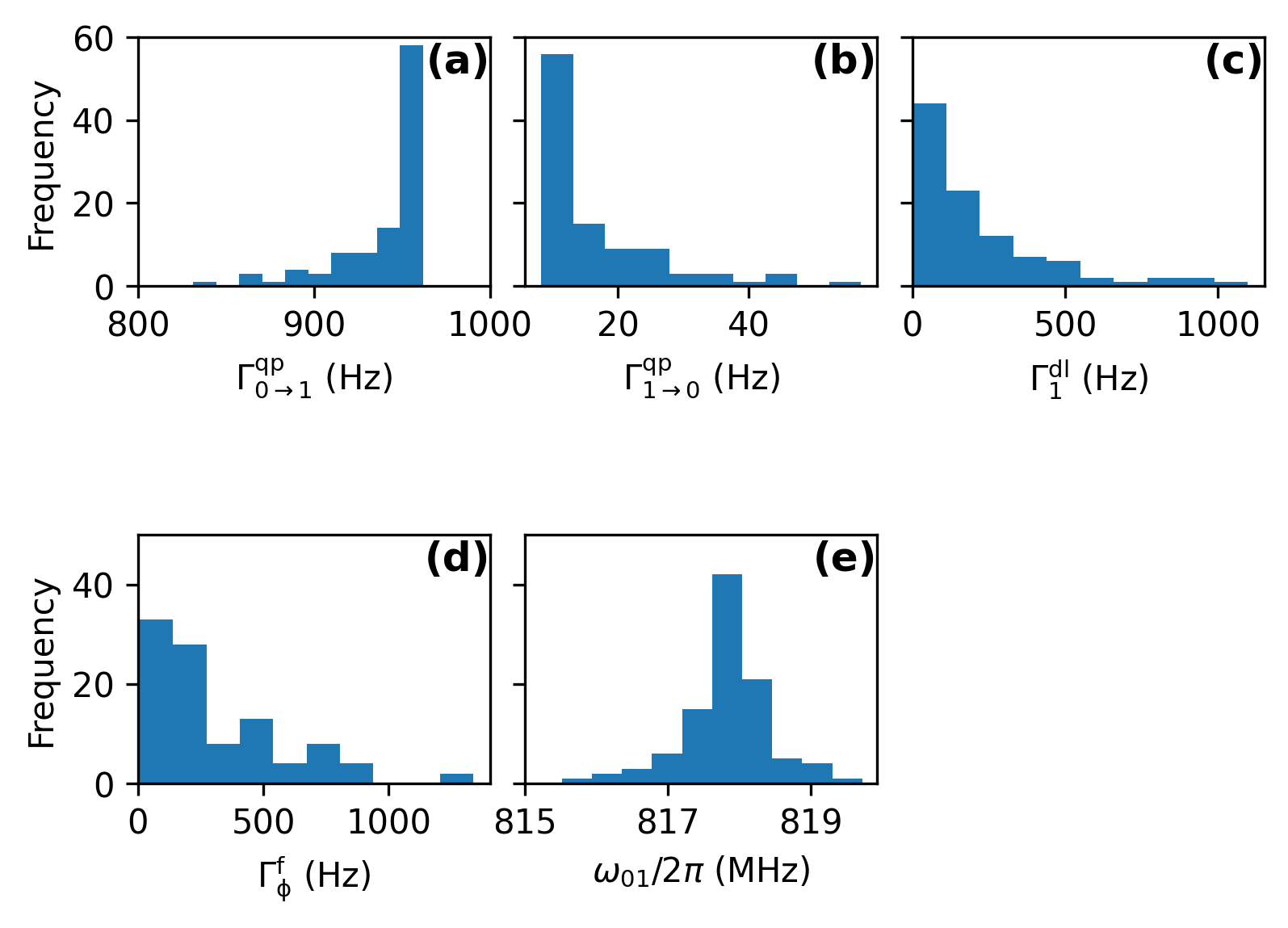}
    \caption{Here, 0.2\% disorder is added to the loop areas. Histograms of (a) excitation and (b) relaxation rates from to quasiparticle tunnelings $T_{0\rightarrow1 }^{\text{qp}} =941$ Hz and $T_{1\rightarrow 0}^{\text{qp}} =16$ Hz, (c) relaxation rate contributed by dielectric loss, with an increased average of 214 Hz, (d) pure dephasing rate due to flux noise, with a slightly increased mean of 312 Hz. This reduces the corresponding pure dephasing time to 3.2 ms. All the other decoherence rates are not affected. (e) Transition frequency $\omega_{01}$, which average is unchanged.}
    \label{fig_e2}
\end{figure*}

\begin{figure*}
    \centering
    \includegraphics[width=0.5\linewidth]{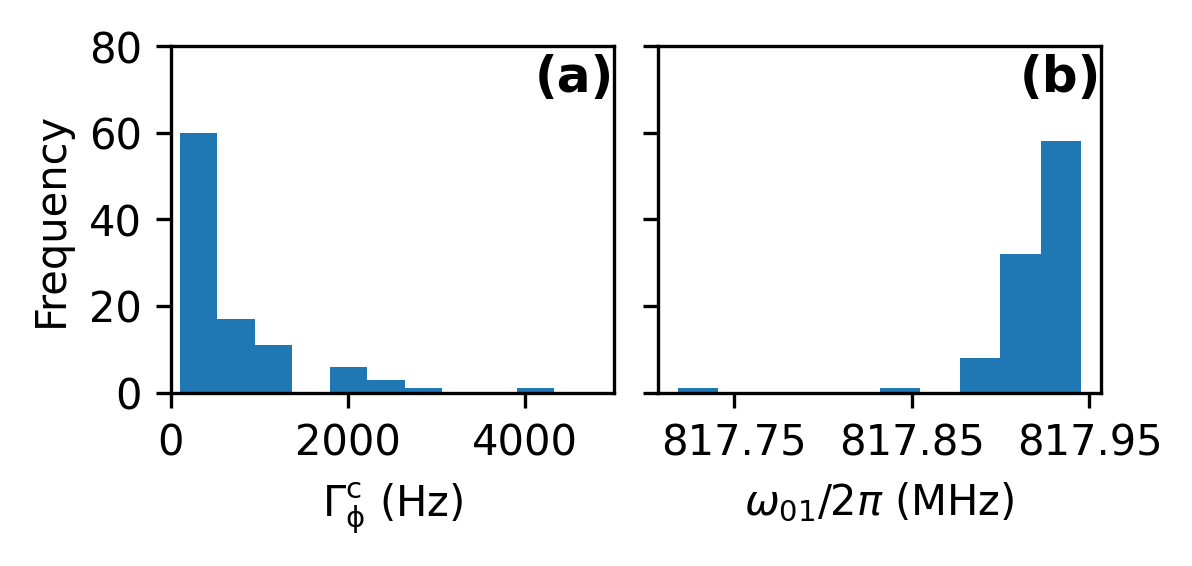}
    \caption{Histograms of (a) charge noise pure dephasing rate whose average is $\Gamma_{\upphi}^{\text{c}}$ = 800 Hz and (b) qubit transition frequency $\omega_{01}$ which is not affected when 0.1\% variation is added to the gate charge.}
    \label{fig_e3}
\end{figure*}

\begin{figure*}
    \centering
    \includegraphics[width=0.7\linewidth]{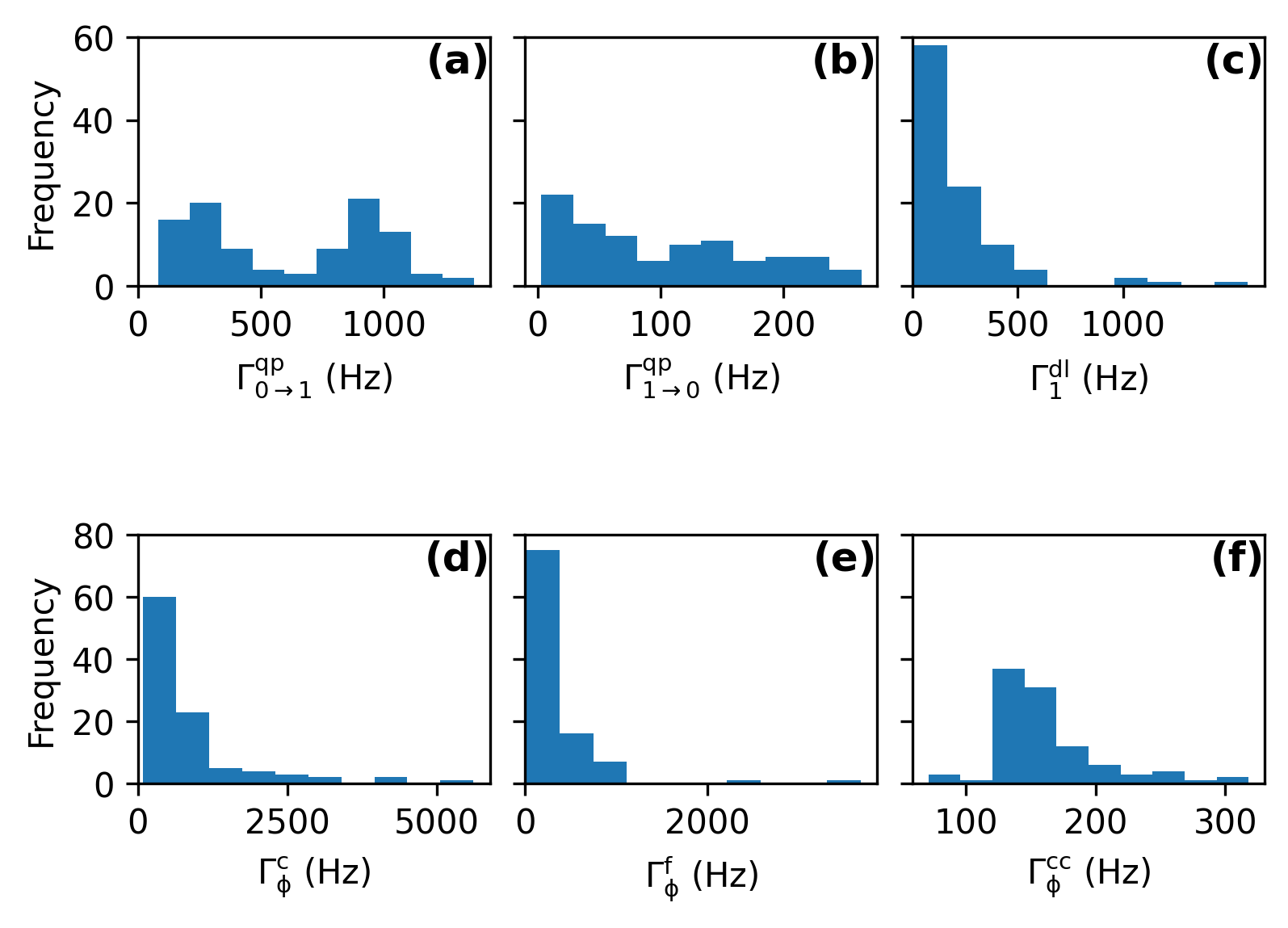}
    \caption{2\% disorder in Josephson junctions, 0.2\% disorder in loop areas, and 0.1\% disorder in gate charges are introduced to the qubit. The average values of the resulting (a) quasiparticle tunneling excitation rate $\Gamma_{0\rightarrow1}^{\text{qp}} = $ 613 Hz, (b) relaxation rate $\Gamma_{1\rightarrow0}^{\text{qp}} = $ 102 Hz, (c) dielectric loss relaxation rate $\Gamma_1^{\text{dl}}$ = 211 Hz, (d) charge noise pure dephasing rate $\Gamma_{\upphi}^{\text{c}}$ = 802 Hz, (e) flux noise pure dephasing rate $\Gamma_{\upphi}^{\text{f}}$ = 329 Hz, (f) critical current noise pure dephasing rate $\Gamma_{\upphi}^{\text{cc}}$ = 162 Hz.}
    \label{fig_e4}
\end{figure*}

To study the individual effects of disorder in junction parameters, loop areas, and gate charges on the qubit’s coherence, we introduce these variations into the qubit parameters separately. First, we add 2\% disorder to the sizes and oxidation parameters of the Josephson junctions and plot the histograms of the resulting variations in Fig. \ref{fig_e1}. We find that under this type of disorder, the qubit's protection is well preserved. In Fig. \ref{fig_e2}, we introduce 0.2\% disorder into the loop areas and plot the histogram of the resulting variations. The disorder in loop areas significantly reduces the dielectric loss relaxation time $T_1^{\text{dl}}$ and flux noise pure dephasing time $T_{\upphi}^{\text{f}}$ but still maintains them at the millisecond range. Next, we apply 0.1\% disorder to the gate charges of the device. It only affects the charge noise pure dephasing time $T_{\upphi}^{\text{c}}$. We plot the histogram of $T_{\upphi}^{\text{c}}$ and $\omega_{01}$ in Fig. \ref{fig_e3} and find the $T_{\upphi}^{\text{c}}$ still preserved at millisecond range. Finally, we study the most realistic case where every type of disorder mentioned above is added to the qubit and plot the resulting variations in Fig. \ref{fig_e4}. We find all the $T_1$ and $T_{\upphi}$ still kept at several milliseconds.

\section{Qubit operation}

\begin{figure*}
    \centering
    \includegraphics[width=\linewidth]{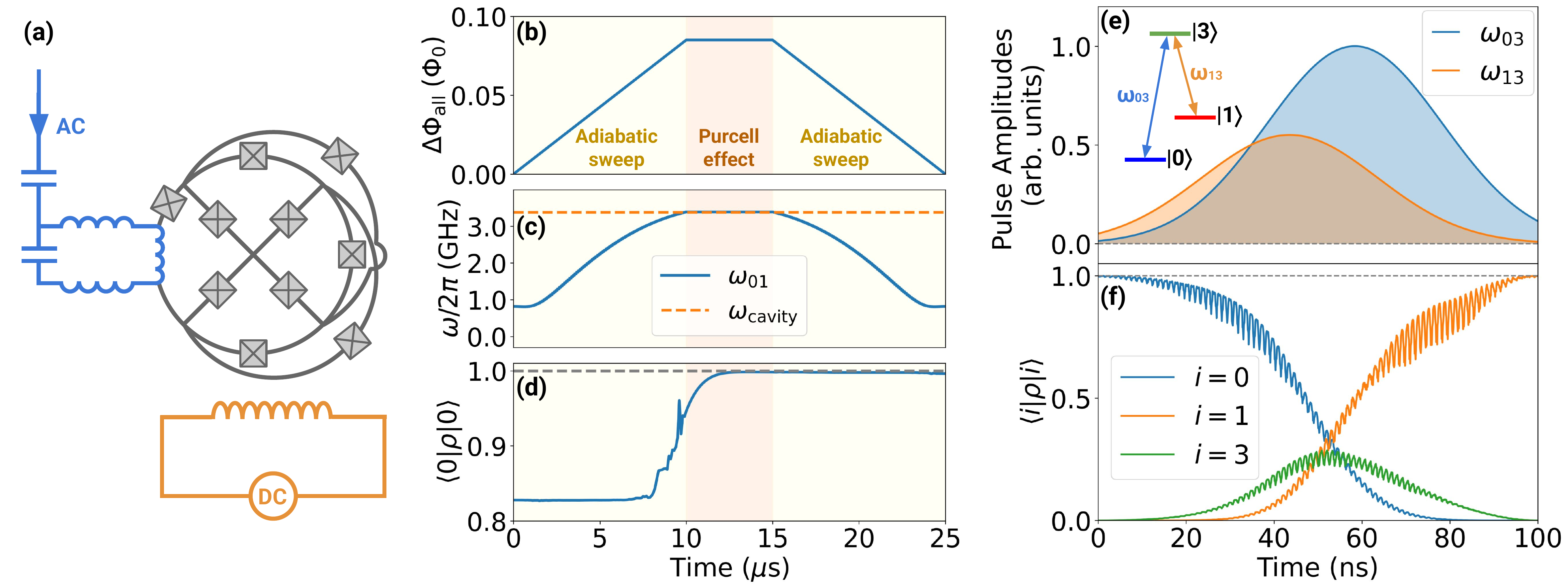}
    \caption{(a) A schematic depicts an operational mode of the qubit. A resonator with a frequency of $\omega_r/2\pi$ = 3.38 GHz and a quality factor of Q = $\omega_r / \kappa$ = 11300 (blue) is connected to the qubit galvanically via a mutual inductance l = 40 pH. Additionally, a DC flux line (orange) is used to bias the qubit, enabling the detuning of the flux threading the qubit from its optimal point for effective initialization. (b) The magnetic flux threading each loop of the qubit is adiabatically tuned to $\Delta \Phi = 1.17 \;\Phi_{opt}$ to match the resonator's frequency. The process takes 10 $\mu s$. The flux is then fixed for 5 $\mu s$ until the coupled system reaches equilibrium, after which the flux is tuned back to the optimal point in 10 $\mu s$. (c) The transition frequency of the qubit, and (d) the fidelity of the ground state during the initialization process. (e) The pulse sequence applied to the qubit for the occupation transfer from $|0\rangle$ to $|1\rangle$. The diagram on the top left indicates the $\Lambda$ system involving $|3\rangle$ state used in the STIRAP procedure. (f) Population of the states $|0\rangle$, $|1\rangle$, and $|3\rangle$ during the STIRAP procedure.}
    \label{fig_f1}
\end{figure*}

Superconducting qubits can be operated by coupling them, either inductively or capacitively, to the control units \cite{krantz2019quantum}. The coupling strengths are proportional to the transition matrix elements $|\langle 0| \hat{I}| 1\rangle|$ and $|\langle 0|\hat{N}| 1\rangle|$ for inductive couplings and capacitive couplings, respectively, where $\hat{I}$ is the junction current operator and $\hat{N}$ is the charge operator. At the optimal point, we calculate $|\langle 0|\hat{I}|1\rangle|$ for all three types of Josephson junctions in the new design and find all of them to be zero. We also find the transition matrix elements of the charge operators $|\langle 0|\hat{N}| 1\rangle| = 0$ for all the sites. It indicates that controlling the qubit via direction transitions between the states is not readily achievable.

When the device's parameters are tuned away from the optimal point, the matrix elements $|\langle 0| \hat{I}| 1\rangle|$ or $|\langle 0|\hat{N}| 1\rangle|$ can become non-zero. We propose a coupling scheme in Fig. \hyperref[fig_f1]{\ref*{fig_f1}(a)} that is similar to the one presented in \cite{brookes2022protection}, where the new qubit is galvanically connected to a quantum LC resonator via a shared inductance $l = 40$ pH that has negligible effects on the symmetry protection of the qubit (see the final paragraph of this section). The resonator has an inductance L = 1.50 nH and a capacitor C = 1.48 pF with a decay rate $\kappa/2\pi = 0.3 $ MHz. The resonator frequency is given by $\omega_{r} = 1/\sqrt{LC}$ = 3.38 GHz. The coupling between the qubit and resonator takes the form
\begin{equation}\label{interaction_hamiltonian}
    H_{int} = l I_{\text{ZPF}}(\hat{a}+\hat{a}^{\dagger}) \hat{I}_a(\Phi_{int},\Phi_{ext}),
\end{equation}
where $I_{\text{ZPF}} = \omega_r\sqrt{\hbar/2Z}$ is the zero point fluctuation of the current in the resonator and $Z = \sqrt{L/C}$ is the impedance of the resonator. $\hat{a}$ and $\hat{a}^{\dagger}$ are the annihilation and creation operators of the resonator. $\hat{I}_a(\Phi_{i})$ is the current operator describing the current flowing in the azimuthal junction next to the shared inductance. The Purcell effect allows fast qubit initialization via the decay of the cavity by coupling the qubit to a cavity and matching their frequencies. When our qubit is moved away from its optimal point to match the resonant frequency of the resonator, we find the transition matrix elements of $\hat{I}_a(\Phi_{i})$ to be non-zero, which enables the coupling between the qubit and the resonator. We numerically simulate the initialization process in a truncated basis consisting of the qubit states $|0\rangle$ and $|1\rangle$. Assuming the qubit sits in a dilution refrigerator cooled to 25 mK, we estimate an initial ground state occupation of 82.8\% using the partition function and a 99.6\% initialization fidelity after a 25 $\mu$s initialization procedure shown in Fig. \hyperref[fig_f1]{\ref*{fig_f1}(b-d)}. Our simulation excludes the higher energy states of the device, implying that, in practice, more intricate initialization protocols may be needed to reduce occupation in higher energy levels.  

Moving the qubit away from its optimal point can enable direct transitions between the qubit states but will leave the qubit unprotected and introduce additional noises. At the optimal point, the transitions between $|0\rangle$ and $|1\rangle$ can be realized using Stimulated Raman adiabatic passage (STIRAP) via higher energy states \cite{kumar2016stimulated}. The advantage of using STIRAP is that the population transfer is immune against the decay of the intermediate state \cite{vitanov2017stimulated}, which is not designed to be intrinsically protected for our qubit. In Fig. \hyperref[fig_f1]{\ref*{fig_f1}(e)}, we propose a STIRAP scheme for our qubit using a $\Lambda$ system involving the $|3\rangle$ state of the device, where both the pump and Stokes drive pulses are resonant with the corresponding transition frequencies $\omega_{03}$ and $\omega_{13}$, respectively. Two Gaussian-shaped pulses with $\sigma = 20$ ns and different peak drive powers $P_{\omega_{04},max}$ = -84 dBm and $P_{\omega_{14},max}$ = -90 dBm are applied to the resonator to make the peak Rabi frequencies nearly equal:
$|\Omega_{04,max}| \approx |\Omega_{14,max}|$. The pulses are applied with a time delay $\tau$ = 15 ns to achieve the optimal adiabaticity \cite{vitanov1997analytic}. Assuming the qubit relaxation rate $\Gamma_1$ = 20 Hz, in our simulation, we use a truncated basis of the lowest five energy states and apply quantum optimal control algorithms in QuTiP \cite{JOHANSSON20131234,johansson2012qutip}. Our numerical simulation demonstrates a population transfer efficiency of 99.9993\% (Fig. \hyperref[fig_f1]{\ref*{fig_f1}(f)}).

Dispersive readout of the qubit state can also be realized using couplings between the qubit states and the device's higher energy states. In the dispersive regime defined by $|\Delta_{ij}| \gg |g_{ij}|\sqrt{1+\bar{n}}$, where $\Delta_{ij} = \omega_i - \omega_j - \omega_r$, $g_{ij} = l I_{ZPF} \langle i|\hat{I}_a(\Phi) |j\rangle$ and $\bar{n}$ is the average photon number in the resonator, the Hamiltonian of the qubit-resonator system in the truncated basis with m states for the qubit device can be written as
\begin{equation}
    \begin{aligned}
    H &= \sum_{i}^m (\omega_i + \Lambda_i)|i\rangle\langle i| + (\omega_r + \sum_i^m\chi_i |i\rangle \langle i|) \hat{a}^{\dagger}\hat{a} \\
    &\simeq \frac{\tilde{\omega}_{01}}{2}\sigma_z + (\tilde{\omega}_r + \chi\sigma_z)\hat{a}^{\dagger}\hat{a},
    \end{aligned}
\end{equation}
where the dispersive shift $\chi_i = \sum_j^m (|g_{ij}|^2/\Delta_{ij}-|g_{ji}|^2/\Delta_{ji})$, the Lamb-shift $\Lambda_i = \sum_j^m |g_{ij}|^2/\Delta_{ij}$, $\tilde{\omega}_{01} = \omega_1-\omega_0 +\Lambda_1 - \Lambda_0$, $\tilde{\omega}_r = \omega_r + (\chi_0 + \chi_1)/2$ and $\chi = (\chi_1 - \chi_0)/2$ \cite{paolo2019control}. At the optimal point, we find that the qubit is located in the dispersive regime and the dispersive shift $\chi/2\pi = - 2.85 $ MHz, whose amplitude is significantly greater than the cavity linewidth $\kappa$, suggesting that the dispersive shifts are large enough to detect the qubit's state.

The effect of introducing a small inductance in the circuit can be equivalently addressed by adjusting the inductance of the Josephson junctions through the following transformation \cite{chang2023tunable}:
\begin{equation}
    L_{J_a} \rightarrow L_{J_a} + l = L_{J_a}(1+\eta),
\end{equation}
where $\eta = l/L_{J_a}$. For our new  qubit, we find $\eta$ = 40 pH/ 33 nH $\approx$ 0.001, which only modifies the Josephson energy $E_{J_a}$ by $\sim$ 6 MHz (0.1\% $E_{J_a}$) and the critical current by $\sim$ 0.01 nA. The alteration to the junction parameters is minor relative to the assumed 2\% disorder in junction parameters, indicating that the new element is unlikely to affect the qubit's symmetry protection.

\bibliographystyle{quantum}
\bibliography{main}

\begin{thebibliography}{10}

\bibitem{cai2023quantum}
Zhenyu Cai, Ryan Babbush, Simon~C Benjamin, Suguru Endo, William~J Huggins, Ying Li, Jarrod~R McClean, and Thomas~E O’Brien.
\newblock ``Quantum error mitigation''.
\newblock \href{https://dx.doi.org/https://doi.org/10.1103/RevModPhys.95.045005}{Reviews of Modern Physics {\bf 95}, 045005}~(2023).

\bibitem{sivak2023real}
VV~Sivak, Alec Eickbusch, Baptiste Royer, Shraddha Singh, Ioannis Tsioutsios, Suhas Ganjam, Alessandro Miano, BL~Brock, AZ~Ding, Luigi Frunzio, et~al.
\newblock ``Real-time quantum error correction beyond break-even''.
\newblock \href{https://dx.doi.org/https://doi.org/10.1038/s41586-023-05782-6}{Nature {\bf 616}, 50--55}~(2023).

\bibitem{google2025quantum}
Google~Quantum AI and Collaborators.
\newblock ``Quantum error correction below the surface code threshold''.
\newblock \href{https://dx.doi.org/https://doi.org/10.1038/s41586-024-08449-y}{Nature {\bf 638}, 920--926}~(2025).

\bibitem{murray2021material}
Conal~E Murray.
\newblock ``Material matters in superconducting qubits''.
\newblock \href{https://dx.doi.org/https://doi.org/10.1016/j.mser.2021.100646}{Materials Science and Engineering: R: Reports {\bf 146}, 100646}~(2021).

\bibitem{premkumar2021microscopic}
Anjali Premkumar, Conan Weiland, Sooyeon Hwang, Berthold J{\"a}ck, Alexander~PM Place, Iradwikanari Waluyo, Adrian Hunt, Valentina Bisogni, Jonathan Pelliciari, Andi Barbour, et~al.
\newblock ``Microscopic relaxation channels in materials for superconducting qubits''.
\newblock \href{https://dx.doi.org/https://doi.org/10.1038/s43246-021-00174-7}{Communications Materials {\bf 2}, 72}~(2021).

\bibitem{chang2022reproducibility}
T~Chang, I~Holzman, T~Cohen, BC~Johnson, DN~Jamieson, and M~Stern.
\newblock ``Reproducibility and gap control of superconducting flux qubits''.
\newblock \href{https://dx.doi.org/https://doi.org/10.1103/PhysRevApplied.18.064062}{Physical Review Applied {\bf 18}, 064062}~(2022).

\bibitem{osman2021simplified}
A~Osman, J~Simon, A~Bengtsson, S~Kosen, P~Krantz, D~P~Lozano, M~Scigliuzzo, P~Delsing, Jonas Bylander, and A~Fadavi~Roudsari.
\newblock ``Simplified josephson-junction fabrication process for reproducibly high-performance superconducting qubits''.
\newblock \href{https://dx.doi.org/https://doi.org/10.1063/5.0037093}{Applied Physics Letters{\bf 118}}~(2021).

\bibitem{groszkowski2018coherence}
Peter Groszkowski, A~Di Paolo, AL~Grimsmo, A~Blais, DI~Schuster, Andrew~Addison Houck, and Jens Koch.
\newblock ``Coherence properties of the 0-$\pi$qubit''.
\newblock \href{https://dx.doi.org/https://doi.org/10.1088/1367-2630/aab7cd}{New Journal of Physics {\bf 20}, 043053}~(2018).

\bibitem{gyenis2021moving}
Andr{\'a}s Gyenis, Agustin Di~Paolo, Jens Koch, Alexandre Blais, Andrew~A Houck, and David~I Schuster.
\newblock ``Moving beyond the transmon: Noise-protected superconducting quantum circuits''.
\newblock \href{https://dx.doi.org/https://doi.org/10.1103/PRXQuantum.2.030101}{PRX Quantum {\bf 2}, 030101}~(2021).

\bibitem{majumdar1969next}
Chanchal~K Majumdar and Dipan~K Ghosh.
\newblock ``On next-nearest-neighbor interaction in linear chain. i''.
\newblock \href{https://dx.doi.org/https://doi.org/10.1063/1.1664978}{Journal of Mathematical Physics {\bf 10}, 1388--1398}~(1969).

\bibitem{levitov2001quantum}
LS~Levitov, TP~Orlando, JB~Majer, and JE~Mooij.
\newblock ``Quantum spin chains and majorana states in arrays of coupled qubits''~(2001).
\newblock  \href{http://arxiv.org/abs/cond-mat/0108266}{arXiv:cond-mat/0108266}.

\bibitem{callison2017protected}
Adam Callison, Eytan Grosfeld, and Eran Ginossar.
\newblock ``Protected ground states in short chains of coupled spins in circuit quantum electrodynamics''.
\newblock \href{https://dx.doi.org/https://doi.org/10.1103/PhysRevB.96.085121}{Physical Review B {\bf 96}, 085121}~(2017).

\bibitem{lidar1998decoherence}
Daniel~A Lidar, Isaac~L Chuang, and K~Birgitta Whaley.
\newblock ``Decoherence-free subspaces for quantum computation''.
\newblock \href{https://dx.doi.org/https://doi.org/10.1103/PhysRevLett.81.2594}{Physical Review Letters {\bf 81}, 2594}~(1998).

\bibitem{lidar2003decoherence}
Daniel~A Lidar and K~Birgitta~Whaley.
\newblock ``Decoherence-free subspaces and subsystems''.
\newblock In Irreversible quantum dynamics.
\newblock \href{https://dx.doi.org/https://doi.org/10.48550/arXiv.quant-ph/0301032}{Pages 83--120}.
\newblock Springer~(2003).

\bibitem{quiroz2024dynamically}
Gregory Quiroz, Bibek Pokharel, Joseph Boen, Lina Tewala, Vinay Tripathi, Devon Williams, Lian-Ao Wu, Paraj Titum, Kevin Schultz, and Daniel Lidar.
\newblock ``Dynamically generated decoherence-free subspaces and subsystems on superconducting qubits''.
\newblock \href{https://dx.doi.org/https://doi.org/10.1088/1361-6633/ad6805}{Reports on Progress in Physics {\bf 87}, 097601}~(2024).

\bibitem{brookes2022protection}
Paul Brookes, Tikai Chang, Marzena Szymanska, Eytan Grosfeld, Eran Ginossar, and Michael Stern.
\newblock ``Protection of quantum information in a chain of josephson junctions''.
\newblock \href{https://dx.doi.org/https://doi.org/10.1103/PhysRevApplied.17.024057}{Physical Review Applied {\bf 17}, 024057}~(2022).

\bibitem{rasmussen2021superconducting}
Stig~Elkj{\ae}r Rasmussen, Kasper~Sangild Christensen, Simon~Panyella Pedersen, Lasse~Bj{\o}rn Kristensen, Thomas B{\ae}kkegaard, Niels Jakob~S{\o}e Loft, and Nikolaj~Thomas Zinner.
\newblock ``Superconducting circuit companion—an introduction with worked examples''.
\newblock \href{https://dx.doi.org/https://doi.org/10.1103/PRXQuantum.2.040204}{PRX Quantum {\bf 2}, 040204}~(2021).

\bibitem{schofield2025roadmap}
Steven~R Schofield, Andrew~J Fisher, Eran Ginossar, Joseph~W Lyding, Richard~M Silver, Fan Fei, Pradeep Namboodiri, Jonathan Wyrick, Mateus~Gallucci Masteghin, David~C Cox, et~al.
\newblock ``Roadmap on atomic-scale semiconductor devices''.
\newblock \href{https://dx.doi.org/https://doi.org/10.1088/2399-1984/ada901}{Nano Futures}~(2025).

\bibitem{singh2022dual}
Kevin Singh, Shraddha Anand, Andrew Pocklington, Jordan~T Kemp, and Hannes Bernien.
\newblock ``Dual-element, two-dimensional atom array with continuous-mode operation''.
\newblock \href{https://dx.doi.org/https://doi.org/10.1103/PhysRevX.12.011040}{Physical Review X {\bf 12}, 011040}~(2022).

\bibitem{groh2016robustness}
Thorsten Groh, Stefan Brakhane, Wolfgang Alt, Dieter Meschede, Janos~K Asb{\'o}th, and Andrea Alberti.
\newblock ``Robustness of topologically protected edge states in quantum walk experiments with neutral atoms''.
\newblock \href{https://dx.doi.org/https://doi.org/10.1103/PhysRevA.94.013620}{Physical Review A {\bf 94}, 013620}~(2016).

\bibitem{scholl2021quantum}
Pascal Scholl, Michael Schuler, Hannah~J Williams, Alexander~A Eberharter, Daniel Barredo, Kai-Niklas Schymik, Vincent Lienhard, Louis-Paul Henry, Thomas~C Lang, Thierry Lahaye, et~al.
\newblock ``Quantum simulation of 2d antiferromagnets with hundreds of rydberg atoms''.
\newblock \href{https://dx.doi.org/https://doi.org/10.1038/s41586-021-03585-1}{Nature {\bf 595}, 233--238}~(2021).

\bibitem{guo2024site}
S-A Guo, Y-K Wu, J~Ye, L~Zhang, W-Q Lian, R~Yao, Y~Wang, R-Y Yan, Y-J Yi, Y-L Xu, et~al.
\newblock ``A site-resolved two-dimensional quantum simulator with hundreds of trapped ions''.
\newblock \href{https://dx.doi.org/https://doi.org/10.1038/s41586-024-07459-0}{NaturePages 1--6}~(2024).

\bibitem{bruzewicz2019trapped}
Colin~D Bruzewicz, John Chiaverini, Robert McConnell, and Jeremy~M Sage.
\newblock ``Trapped-ion quantum computing: Progress and challenges''.
\newblock \href{https://dx.doi.org/https://doi.org/10.1063/1.5088164}{Applied Physics Reviews{\bf 6}}~(2019).

\bibitem{britton2012engineered}
Joseph~W Britton, Brian~C Sawyer, Adam~C Keith, C-C~Joseph Wang, James~K Freericks, Hermann Uys, Michael~J Biercuk, and John~J Bollinger.
\newblock ``Engineered two-dimensional ising interactions in a trapped-ion quantum simulator with hundreds of spins''.
\newblock \href{https://dx.doi.org/https://doi.org/10.1038/nature10981}{Nature {\bf 484}, 489--492}~(2012).

\bibitem{chen2016measuring}
Zijun Chen, Julian Kelly, Chris Quintana, R~Barends, B~Campbell, Yu~Chen, B~Chiaro, A~Dunsworth, AG~Fowler, E~Lucero, et~al.
\newblock ``Measuring and suppressing quantum state leakage in a superconducting qubit''.
\newblock \href{https://dx.doi.org/https://doi.org/10.1103/PhysRevLett.116.020501}{Physical review letters {\bf 116}, 020501}~(2016).

\bibitem{werninghaus2021leakage}
Max Werninghaus, Daniel~J Egger, Federico Roy, Shai Machnes, Frank~K Wilhelm, and Stefan Filipp.
\newblock ``Leakage reduction in fast superconducting qubit gates via optimal control''.
\newblock \href{https://dx.doi.org/https://doi.org/10.1038/s41534-020-00346-2}{npj Quantum Information {\bf 7}, 14}~(2021).

\bibitem{mcewen2021removing}
Matt McEwen, Dvir Kafri, Z~Chen, Juan Atalaya, KJ~Satzinger, Chris Quintana, Paul~Victor Klimov, Daniel Sank, C~Gidney, AG~Fowler, et~al.
\newblock ``Removing leakage-induced correlated errors in superconducting quantum error correction''.
\newblock \href{https://dx.doi.org/https://doi.org/10.1038/s41467-021-21982-y}{Nature communications {\bf 12}, 1761}~(2021).

\bibitem{terhal2015quantum}
Barbara~M Terhal.
\newblock ``Quantum error correction for quantum memories''.
\newblock \href{https://dx.doi.org/https://doi.org/10.1103/RevModPhys.87.307}{Reviews of Modern Physics {\bf 87}, 307--346}~(2015).

\bibitem{qiu2014coupling}
Yueyin Qiu, Wei Xiong, Lin Tian, and JQ~You.
\newblock ``Coupling spin ensembles via superconducting flux qubits''.
\newblock \href{https://dx.doi.org/https://doi.org/10.1103/PhysRevA.89.042321}{Physical Review A {\bf 89}, 042321}~(2014).

\bibitem{pita2025blueprint}
Marta Pita-Vidal, Jaap~J Wesdorp, and Christian~Kraglund Andersen.
\newblock ``Blueprint for all-to-all-connected superconducting spin qubits''.
\newblock \href{https://dx.doi.org/https://doi.org/10.1103/PRXQuantum.6.010308}{PRX Quantum {\bf 6}, 010308}~(2025).

\bibitem{wang2017observing}
Xin Wang, Adam Miranowicz, Hong-Rong Li, and Franco Nori.
\newblock ``Observing pure effects of counter-rotating terms without ultrastrong coupling: A single photon can simultaneously excite two qubits''.
\newblock \href{https://dx.doi.org/https://doi.org/10.1103/PhysRevA.96.063820}{Physical Review A {\bf 96}, 063820}~(2017).

\bibitem{hita2022ultrastrong}
Mar{\'\i}a Hita-P{\'e}rez, Gabriel Jaum{\`a}, Manuel Pino, and Juan~Jos{\'e} Garc{\'\i}a-Ripoll.
\newblock ``Ultrastrong capacitive coupling of flux qubits''.
\newblock \href{https://dx.doi.org/https://doi.org/10.1103/PhysRevApplied.17.014028}{Physical Review Applied {\bf 17}, 014028}~(2022).

\bibitem{kusunoki2002dielectric}
M~Kusunoki, M~Inadomaru, S~Ohshima, K~Aizawa, M~Mukaida, M~Lorenz, and H~Hochmuth.
\newblock ``Dielectric loss tangent of sapphire single crystal produced by edge-defined film-fed growth method''.
\newblock \href{https://dx.doi.org/https://doi.org/10.1016/S0921-4534(01)01282-5}{Physica C: Superconductivity {\bf 377}, 313--318}~(2002).

\bibitem{read2023precision}
Alexander~P Read, Benjamin~J Chapman, Chan~U Lei, Jacob~C Curtis, Suhas Ganjam, Lev Krayzman, Luigi Frunzio, and Robert~J Schoelkopf.
\newblock ``Precision measurement of the microwave dielectric loss of sapphire in the quantum regime with parts-per-billion sensitivity''.
\newblock \href{https://dx.doi.org/https://doi.org/10.1103/PhysRevApplied.19.034064}{Physical Review Applied {\bf 19}, 034064}~(2023).

\bibitem{deng2023titanium}
Hao Deng, Zhijun Song, Ran Gao, Tian Xia, Feng Bao, Xun Jiang, Hsiang-Sheng Ku, Zhisheng Li, Xizheng Ma, Jin Qin, et~al.
\newblock ``Titanium nitride film on sapphire substrate with low dielectric loss for superconducting qubits''.
\newblock \href{https://dx.doi.org/https://doi.org/10.1103/PhysRevApplied.19.024013}{Physical Review Applied {\bf 19}, 024013}~(2023).

\bibitem{kamenov2023suppression}
Plamen Kamenov, Thomas DiNapoli, Michael Gershenson, and Srivatsan Chakram.
\newblock ``Suppression of quasiparticle poisoning in transmon qubits by gap engineering''~(2023).
\newblock  \href{http://arxiv.org/abs/2309.02655}{arXiv:2309.02655}.

\bibitem{diamond2022distinguishing}
Spencer Diamond, Valla Fatemi, Max Hays, Heekun Nho, Pavel~D Kurilovich, Thomas Connolly, Vidul~R Joshi, Kyle Serniak, Luigi Frunzio, Leonid~I Glazman, et~al.
\newblock ``Distinguishing parity-switching mechanisms in a superconducting qubit''.
\newblock \href{https://dx.doi.org/https://doi.org/10.1103/PRXQuantum.3.040304}{PRX Quantum {\bf 3}, 040304}~(2022).

\bibitem{pan2022engineering}
Xianchuang Pan, Yuxuan Zhou, Haolan Yuan, Lifu Nie, Weiwei Wei, Libo Zhang, Jian Li, Song Liu, Zhi~Hao Jiang, Gianluigi Catelani, et~al.
\newblock ``Engineering superconducting qubits to reduce quasiparticles and charge noise''.
\newblock \href{https://dx.doi.org/https://doi.org/10.1038/s41467-022-34727-2}{Nature Communications {\bf 13}, 7196}~(2022).

\bibitem{gordon2022environmental}
RT~Gordon, Conal~E Murray, C~Kurter, M~Sandberg, SA~Hall, Karthik Balakrishnan, R~Shelby, B~Wacaser, AA~Stabile, JW~Sleight, et~al.
\newblock ``Environmental radiation impact on lifetimes and quasiparticle tunneling rates of fixed-frequency transmon qubits''.
\newblock \href{https://dx.doi.org/https://doi.org/10.1063/5.0078785}{Applied Physics Letters{\bf 120}}~(2022).

\bibitem{diamond2024quasiparticles}
Spencer Diamond.
\newblock ``Quasiparticles and charge-parity switching in transmon qubits''.
\newblock \href{https://dx.doi.org/https://www.proquest.com/dissertations-theses/quasiparticles-charge-parity-switching-transmon/docview/3084696514/se-2?accountid=14511}{PhD thesis}.
\newblock Yale University.
\newblock ~(2024).

\bibitem{chang2023tunable}
T~Chang, T~Cohen, I~Holzman, G~Catelani, and M~Stern.
\newblock ``Tunable superconducting flux qubits with long coherence times''.
\newblock \href{https://dx.doi.org/https://doi.org/10.1103/PhysRevApplied.19.024066}{Physical Review Applied {\bf 19}, 024066}~(2023).

\bibitem{krantz2019quantum}
Philip Krantz, Morten Kjaergaard, Fei Yan, Terry~P Orlando, Simon Gustavsson, and William~D Oliver.
\newblock ``A quantum engineer's guide to superconducting qubits''.
\newblock \href{https://dx.doi.org/https://doi.org/10.1063/1.5089550}{Applied physics reviews{\bf 6}}~(2019).

\bibitem{PRIMME}
Andreas Stathopoulos and James~R. McCombs.
\newblock ``{PRIMME}: {PR}econditioned {I}terative {M}ulti{M}ethod {E}igensolver: Methods and software description''.
\newblock \href{https://dx.doi.org/https://doi.org/10.1145/1731022.1731031}{ACM Transactions on Mathematical Software {\bf 37}, 21:1--21:30}~(2010).

\bibitem{svds_software}
Lingfei Wu, Eloy Romero, and Andreas Stathopoulos.
\newblock ``Primme{\_}svds: {A} high-performance preconditioned {SVD} solver for accurate large-scale computations''.
\newblock \href{https://dx.doi.org/https://doi.org/10.1137/16M1082214}{SIAM Journal on Scientific Computing {\bf 39}, S248--S271}~(2017).

\bibitem{brookes2021protected}
Paul Brookes.
\newblock ``Protected states and metastable dynamics in superconducting circuits''.
\newblock \href{https://dx.doi.org/https://discovery.ucl.ac.uk/id/eprint/10130287/}{PhD thesis}.
\newblock UCL (University College London).
\newblock ~(2021).

\bibitem{stern2014flux}
Michael Stern, Gianluigi Catelani, Yuimaru Kubo, Cecile Grezes, Audrey Bienfait, Denis Vion, Daniel Esteve, and Patrice Bertet.
\newblock ``Flux qubits with long coherence times for hybrid quantum circuits''.
\newblock \href{https://dx.doi.org/https://doi.org/10.1103/PhysRevLett.113.123601}{Physical review letters {\bf 113}, 123601}~(2014).

\bibitem{ithier2005manipulation}
Gr{\'e}goire Ithier.
\newblock ``Manipulation, readout and analysis of the decoherence of a superconducting quantum bit''.
\newblock \href{https://dx.doi.org/https://theses.hal.science/tel-00130589/}{PhD thesis}.
\newblock Universit{\'e} Pierre et Marie Curie-Paris VI.
\newblock ~(2005).

\bibitem{kumar2016stimulated}
KS~Kumar, Antti Veps{\"a}l{\"a}inen, S~Danilin, and GS~Paraoanu.
\newblock ``Stimulated raman adiabatic passage in a three-level superconducting circuit''.
\newblock \href{https://dx.doi.org/https://doi.org/10.1038/ncomms10628}{Nature communications {\bf 7}, 10628}~(2016).

\bibitem{vitanov2017stimulated}
Nikolay~V Vitanov, Andon~A Rangelov, Bruce~W Shore, and Klaas Bergmann.
\newblock ``Stimulated raman adiabatic passage in physics, chemistry, and beyond''.
\newblock \href{https://dx.doi.org/https://doi.org/10.1103/RevModPhys.89.015006}{Reviews of Modern Physics {\bf 89}, 015006}~(2017).

\bibitem{vitanov1997analytic}
NV~Vitanov and Stig Stenholm.
\newblock ``Analytic properties and effective two-level problems in stimulated raman adiabatic passage''.
\newblock \href{https://dx.doi.org/https://doi.org/10.1103/PhysRevA.55.648}{Physical Review A {\bf 55}, 648}~(1997).

\bibitem{JOHANSSON20131234}
J.R. Johansson, P.D. Nation, and Franco Nori.
\newblock ``Qutip 2: A python framework for the dynamics of open quantum systems''.
\newblock \href{https://dx.doi.org/https://doi.org/10.1016/j.cpc.2012.11.019}{Computer Physics Communications {\bf 184}, 1234--1240}~(2013).

\bibitem{johansson2012qutip}
J~Robert Johansson, Paul~D Nation, and Franco Nori.
\newblock ``Qutip: An open-source python framework for the dynamics of open quantum systems''.
\newblock \href{https://dx.doi.org/https://doi.org/10.1016/j.cpc.2012.02.021}{Computer physics communications {\bf 183}, 1760--1772}~(2012).

\bibitem{paolo2019control}
Agustin~Di Paolo, Arne~L Grimsmo, Peter Groszkowski, Jens Koch, and Alexandre Blais.
\newblock ``Control and coherence time enhancement of the 0--$\pi$ qubit''.
\newblock \href{https://dx.doi.org/https://doi.org/10.1088/1367-2630/ab09b0}{New Journal of Physics {\bf 21}, 043002}~(2019).

\end{thebibliography}

\end{document}